\documentclass[final,authoryear,5p,times,twocolumn]{elsarticle}

\usepackage{lineno}
\usepackage{amsmath,amssymb}
\usepackage[colorlinks=True,%
            urlcolor=blue,%
            linkcolor=blue,%
            citecolor= blue,%
            pdfauthor={T. Gastine},%
            pdftitle={Zonal flow regimes in rotating anelastic spherical shells:
an application to giant planets},%
            pdfkeywords={Atmospheres dynamics, Jupiter interior, Saturn
interior, Uranus interior, Neptune interior},%
            dvips]{hyperref}
\usepackage{journals} 
\usepackage{booktabs}
\usepackage{longtable}

\journal{Icarus}

\newcommand{\lp}{\left(}
\newcommand{\rp}{\right)}
\newcommand{\tb}{\tilde{T}}
\newcommand{\rb}{\tilde{\rho}}

\newcommand{\vna}{\vec{\nabla}}

\newcommand{\mom}{{\cal L}}

\newcommand{\ras}{{\cal R}^*}
\newcommand{\rasmid}{{\cal R}^*_m}

\def\vec#1{\ensuremath{\mathchoice{\mbox{\boldmath$\displaystyle#1$}}
{\mbox{\boldmath$\textstyle#1$}}
{\mbox{\boldmath$\scriptstyle#1$}}
{\mbox{\boldmath$\scriptscriptstyle#1$}}}}
\def\tens#1{\ensuremath{\mathsf{#1}}}

\begin{document}

\begin{frontmatter}

\title{Zonal flow regimes in rotating anelastic spherical shells: an 
application to giant planets}

\author[MPS]{T.~Gastine\corref{cor1}}
\ead{gastine@mps.mpg.de}
\author[MPS]{J.~Wicht}
\author[UCLA]{J.M.~Aurnou}

\address[MPS]{Max Planck Institut f\"ur Sonnensytemforschung, Max Planck Strasse
2, 37191 Katlenburg Lindau, Germany}
\address[UCLA]{Department of Earth and Space Sciences, University of 
California, Los Angeles, CA, 90095-1567 USA}

\cortext[cor1]{Principal corresponding author}

\begin{abstract}
The surface zonal winds observed in the giant planets form a complex jet
pattern with alternating prograde and retrograde direction. While the main
equatorial band is prograde on the gas giants, both ice giants have a pronounced
retrograde equatorial jet.

We use three-dimensional numerical models of compressible convection
in rotating spherical shells to explore the properties of zonal flows in
different regimes where either rotation or buoyancy dominates the force
balance. We conduct a systematic parameter study to quantify the dependence of
zonal flows on the background density stratification and the driving of
convection.

In our numerical models, we find that the direction of the equatorial zonal wind
is controlled by the ratio of the global-scale buoyancy force and the Coriolis
force. The prograde equatorial band maintained by
Reynolds stresses is found in the rotation-dominated regime.
In cases where buoyancy dominates Coriolis force, the angular momentum per unit
mass is homogenised and the equatorial band is retrograde, reminiscent to
those observed in the ice giants. In this regime, the amplitude of
the zonal jets depends on the background density contrast with strongly
stratified models producing stronger jets than comparable weakly stratified
cases. Furthermore, our results can help to explain the transition between
solar-like (i.e. prograde at the equator) and the ``anti-solar'' differential
rotations (i.e. retrograde at the equator) found in anelastic
models of stellar convection zones.

In the strongly stratified cases, we find that the leading order force balance
can significantly vary with depth. While the flow in the deep interior is
dominated by rotation, buoyancy can indeed become larger than Coriolis force in
a thin region close to the surface. This so-called ``transitional regime'' has a
visible signature in the  main equatorial jet which shows a pronounced dimple
where flow amplitudes notably decay towards the equator. A similar dimple is
observed on Jupiter, which suggests that convection in the planet interior could
possibly operate in this regime.

\end{abstract}


\end{frontmatter}

\section{Introduction}

Surface zonal jets on the giant planets Jupiter, Saturn, Uranus and Neptune
have been investigated since the 1970's by tracking cloud features
\citep[e.g.][]{Ingersoll90}. On each planet, these zonal winds form a
differential rotation profile with alternating prograde (i.e. eastward) and
retrograde (i.e. westward) flows.

On both gas giants, a strong eastward equatorial jet is flanked by
multiple weaker alternating zonal winds (10-20 m.s$^{-1}$). Jupiter's equatorial
jet extends roughly between $\pm20^\circ$ latitude reaching a maximum velocity
around 150 m.s$^{-1}$ \citep{Porco03,Vasavada05}.
Saturn's equatorial jet is fiercer and wider with a maximum flow amplitude of 450
m.s$^{-1}$ and extension of $\pm 35^\circ$ latitude \citep{Porco05}.


The zonal wind profiles of Uranus and Neptune are quite different.
One broad retrograde equatorial jet
is flanked by only two strong prograde jets at higher latitudes. On Uranus, this
equatorial sub-rotation extends between
$\pm 30^\circ$ latitude and reaches 100 m.s$^{-1}$ \citep{Hammel05}; 
it extends over $\pm50^\circ$ latitudes and reaches 400 m.s$^{-1}$ on
Neptune \citep{Sromovsky93}.

Two competing categories of models try to explain the observed zonal winds
structure. In the ``shallow-forcing'' scenario, zonal winds are driven by
injection of turbulence via different types of physical forcings at the
stably-stratified cloud level (e.g., latent heat release, solar radiation, moist
convection). These models successfully reproduce an alternating zonal flow
pattern similar to those observed in giant
planets \citep[e.g.][]{Williams78,Cho96}. Although many earlier shallow models
predict a similar westward equatorial zonal flow for all four giant planets,
recent studies show that these models can also replicate the equatorial zonal
flows of the four giant planets via the inclusion of an additional
forcing mechanism such as water vapor condensation \citep{Lian10}
or via enhanced radiative damping \citep{Scott08,Liu11}.

In the ``deep-forcing'' scenario, zonal winds are maintained by deep-seated
convection. The increasing electrical conductivity in the
deep-interior of the giant planets  \citep{Nettelmann08,French12} goes along
with stronger Lorentz forces that are thought to prevent the strong zonal winds
to penetrate deep into the molecular layer. \cite{Liu08} therefore argued
that the zonal winds must be confined in a thin upper layer ($0.85~R_S$ and
$0.96~R_J$), since deep jets extending over the whole molecular envelope would
lead to Ohmic dissipation that would exceed the planetary luminosity.
More recent dynamo models by \cite{Heimpel11} support the idea that
zonal flows would maybe reach less than half of the distance to the bottom of
the molecular layer. These deep models rely on 3-D numerical simulations of
rapidly-rotating spherical shells. In rotating convection at moderate convective
forcing, convection occurs on long, axially-oriented columns reaching through
the whole fluid layer. These well-organized columnar flows generate Reynolds
stresses (i.e. statistical correlations between the convective flow components)
that drive strong zonal winds \citep[e.g.][]{Busse83,Busse94,Plaut08,
Brown08}.
The typical azimuthally prograde tilt of these convective columns always yields
an eastward equatorial jet \citep[e.g.][]{Zhang92,Christensen01,Aurnou04}. The
direction and the number of jets are consistent with Jupiter's and Saturn's
observation provided thin shells and low Ekman numbers are considered
\citep[e.g.][]{Heimpel05,Heimpel07}.                                           

A different zonal flow regime is however found when global-scale buoyancy starts
to dominate the Coriolis force. The equatorial zonal flow
tends to reverse \citep[e.g.][]{Gilman77,Gilman79,Glatz5}. Responsible for the
latter is the turbulent mixing of angular momentum, which may explain the strong
retrograde equatorial zonal flow observed on the ice giants \citep{Aurnou07}.
%

To further inform the ongoing discussion on the driving mechanisms
and the depth of the zonal jets in the giant planets, here we investigate 
the deep-seated zonal flow perspective and thus compute 3-D global
models of convection in spherical shells.

Many of the previous parameter studies have
employed the Boussinesq approximation where the density stratification is simply
ignored \citep{Christensen02,Aurnou07}. This is rather dubious in the strongly stratified giant planets
where the density contrasts are huge \citep[e.g.][]{Nettelmann12,Nettelmann12a}.
More recent models therefore use the anelastic approximation which allows to
incorporate the effects of background stratification while filtering out the
fast acoustic waves \citep[e.g.][]{Brag95,Lantz99,Brown12}. In an
extensive parameter study, \cite{Gastine12} concentrate on the influence of the
density stratification on convection and zonal flows in the rotation-dominated
regime \citep[see also][]{Showman10}. While the density stratification affects
the local scales and the amplitude of the convective flow, the mean zonal flows
and the global quantities are fairly independent of the density contrast,
similar to the results of \cite{Jones09}. 

Many anelastic and fully compressible models of solar and stellar convection have
observed a transition between the solar-like (i.e. eastward equatorial
zonal flow) and the so-called ``anti-solar'' (i.e. westward equatorial
zonal flow) differential rotation profiles when buoyancy dominates the force
balance \citep{Glatz5,Bessolaz11,Kapyla11}. 
Yet, to date, no systematic parameter study has been made to investigate the
influence of density stratification on the transition between the
rotation-dominated and the buoyancy-dominated zonal flow regimes.

This is precisely the focus of the present study, which extends the previous
Boussinesq study of \cite{Aurnou07} to anelastic models.
To this end, we conduct a systematic parameter study from Boussinesq to
strongly stratified models (i.e. $\rho_\text{bot}/\rho_\text{top} \simeq 150$)
and solutions that span the range from weakly to strongly supercritical
convection.

In section~\ref{sec:model}, we present the anelastic formulation and the numerical
methods. Section~\ref{sec:buo} shows the change in convection when the
driving is gradually increased from the rotation-dominated to the
buoyancy-dominated regime. Section~\ref{sec:amm} focuses on the zonal flows
profiles that develop in the buoyancy-dominated regime.
In section~\ref{sec:dimple}, we concentrate on the so-called \emph{transitional
regime}, a specific feature of strongly stratified anelastic models, before
concluding in section~\ref{sec:conclu}.

\section{Hydrodynamical model and numerical methods}
\label{sec:model}

\subsection{Governing equations}

We consider hydrodynamical simulations of an anelastic ideal gas in a spherical
shell rotating at a constant rotation rate $\Omega$ about the $z$-axis.
We use a dimensionless formulation of the governing Navier-Stokes equations
where the shell thickness $d=r_o-r_i$ is employed as a reference lengthscale and
$\Omega^{-1}$ as the time unit. Density and temperature are non-dimensionalised
using their outer boundaries reference values $\rho_{\text{top}}$ and
$T_{\text{top}}$. Entropy is expressed in units of $\Delta s$, the imposed
entropy contrast over the layer. Kinematic viscosity $\nu$, thermal diffusivity
$\kappa$ and heat capacity $c_p$ are assumed to be homogeneous.

Following the anelastic formulations of \cite{Glatz1,Brag95} and \cite{Lantz99},
the background reference state (denoted with tildes in the following) is
hydrostatic and adiabatic. It is defined by $d\tb/dr = -g/c_p$ and a
polytropic gas $\rb = \tb^m$, $m$ being the polytropic index.
As we are interested in the dynamics of the molecular region of giant planets,
we assume that the mass is concentrated in the inner part, so that
$g \propto 1/r^2$ provides a good approximation \citep[see
also][]{Jones09,Gastine12}. Such a gravity profile then leads to the following
background temperature profile

\begin{equation}
  \tb(r) = \dfrac{c_0}{(1-\eta)r}+ 1-c_0 \quad\text{and}\quad\rb(r) = \tb^m,
\end{equation}
with

\begin{equation}
c_0 = \dfrac{\eta}{1-\eta}\lp \exp\dfrac{N_\rho}{m} -1 \rp \quad\text{with}\quad
N_\rho = \ln\dfrac{\rb(r_i)}{\rb(r_o)}.
\end{equation}
Here $\tb$ and $\rb$  are the background temperature and density,
$\eta=r_i/r_o$ is the aspect ratio of the spherical shell and $N_\rho$
corresponds to the number of density scale heights over the layer
\citep[see also][for the full derivation of the reference state]{Jones11}.

In the anelastic approximation, the dimensionless equations that govern
convective motions are given by

\begin{equation}
 \vec{\nabla}\cdot \lp \rb\vec{u} \rp = 0,
 \label{eq:anel}
\end{equation}
\begin{equation}
   \dfrac{\partial \vec{u}}{\partial
   t}+\vec{u}\cdot\vec{\nabla}\vec{u}
+2\vec{e_z}\times\vec{u}
    =
 -\vec{\nabla}{\dfrac{p}{\rb}}+\text{Ra}^*\dfrac{r_o^2}{r^2} s\,\vec{e_r} 
+  \dfrac{\text{E}}{\rb} \vec{\nabla}\cdot\tens{S},
 \label{eq:NS}
\end{equation}
where $\vec{u}$, $p$  and $s$ are velocity, pressure
and entropy, respectively. $\tens{S}$ is the traceless rate-of-strain tensor
with a constant kinematic viscosity, given by

\begin{equation}
\tens{S}_{ij}= 2 \rb\left(\tens{e}_{ij}- \frac{1}{3} \delta_{ij}
\vec{\nabla}\cdot\vec{u} \right) \quad\text{with}\quad
\tens{e}_{ij} = \dfrac{1}{2}\left(\frac{\partial u_i}{\partial x_j} +
\frac{\partial u_j}{\partial x_i}\right),
\label{eq:tenseur}
\end{equation}
$\delta_{ij}$ being the identity matrix. The dimensionless entropy equation
reads

\begin{equation}
\rb\tb\lp\dfrac{\partial s}{\partial t} + \vec{u}\cdot\vec{\nabla} s\rp =
\dfrac{\text{E}}{\text{Pr}}\vec{\nabla}\cdot\lp\rb\tb \vec{\nabla} s\rp +
\dfrac{\text{E}}{\text{Ra}^*}(1-\eta)c_o\,Q_\nu,
\label{eq:entropy}
\end{equation}
where $Q_\nu$ is the viscous heating contribution given by

\begin{equation}
 Q_\nu =
2\rb\left[\tens{e}_{ij}\tens{e}_{ji}-\dfrac{1}{3}(\vec{\nabla}\cdot\vec{u}
)^2\right ].
\end{equation}
In addition to the aspect ratio $\eta$
and the two parameters involved in the description of the reference state
($N_\rho$ and $m$), the system of equations (\ref{eq:anel}-\ref{eq:entropy}) is
governed by three dimensionless parameters, namely, the Ekman number, the
Prandtl number and the modified Rayleigh number:

\begin{equation}
 \text{E} = \dfrac{\nu}{\Omega d^2} \ ;\ \text{Pr} =
\dfrac{\nu}{\kappa} \ ;\  \text{Ra}^* = \dfrac{g_{\text{top}} \Delta
s}{c_p \Omega^2 d},
\label{eq:params}
\end{equation}
where $g_{\text{top}}$ is the gravity at the outer boundary. $\text{Ra}^*$ can
be related to the standard Rayleigh number $\text{Ra}= g_{\text{top}} d^3 \Delta
s / c_p \nu \kappa$ with \citep[e.g.][]{Christensen02}

\begin{equation}
 \text{Ra}^* = \dfrac{\text{Ra}\, \text{E}^2}{\text{Pr}}.
\label{eq:rastardef}
\end{equation}
The definition of $\text{Ra}^*$ is based on the global entropy jump
over the spherical shell and on the gravity value at the outer
boundary. In anelastic models, it is however more appropriate to define a
local modified Rayleigh number that encompasses the radial dependence of the 
background
reference state \citep[e.g.][]{Kaspi09,Gastine12}:

\begin{equation}
 \ras (r) = \dfrac{g}{c_p \Omega^2}\, \left|\dfrac{d s_c}{dr}\right|
\label{eq:raloc},
\end{equation}
where $s_c$ is the conduction state entropy, which is the solution of

\begin{equation}
  \quad \vec{\nabla}\cdot\lp\rb\tb\vec{\nabla}s_c\rp = 0.
\end{equation}
As the entropy gradient is inversely proportional to $\rb\tb$, the value
of $\ras$ can become very large at the surface in the most stratified cases. 
In
these models, convection sets in first in the outermost region \citep{Jones09a,
Gastine12}. To compare numerical models with different density contrasts, it is
either possible to consider a mass-weighted average of $\ras$ as
suggested by \cite{Kaspi09}, or use its value at mid-depth
\citep{Unno60,Glatz2}. Table~\ref{tab:rascaling} shows that these two
definitions lead to very similar results. In the following, we use the modified
Rayleigh number at mid-depth $\rasmid \equiv \ras(r_{\text{mid}})$
to compare our different numerical models.

\begin{table}
\caption{Local Rayleigh number at mid-depth and mass-weighted average of $\ras$
for $\text{Ra}^*=1$ for different density stratifications.}
\centering
\begin{tabular}{cccc}
  \toprule
  $N_\rho$ & $\ras(r_\text{mid})$ & $\langle \ras \rangle_\rho$ \\
  \midrule
 0.01 & 1.462 & 1.534 \\
 1 & 1.563 & 1.599 \\
 3 & 0.880 & 0.867 \\
 5 & 0.237 & 0.245 \\
\bottomrule
 \end{tabular}
\label{tab:rascaling}
\end{table}

The first numerical models of rotating convection in spherical shells by
\cite{Gilman77} and \cite{Glatz5} have shown that the physical mechanism
responsible of the zonal flow production is sensitive to the relative
contribution of buoyancy and Coriolis force in the global-scale 
force balance. The ratio between these two forces can be related to 
$\text{Ra}^*$ \citep[for the full derivation, see][]{Aurnou07} via

\begin{equation}
 \dfrac{\text{Buoyancy}}{\text{Coriolis}} \sim \lp  \text{Ra}^*  \rp^{1/2}.
\end{equation}
This ratio is commonly referred to as the ``convective Rossby number''
in the solar and stellar convection communities
\citep[e.g.][]{Elliott00,Ballot07} and in the
fluid physics community \citep[e.g.][]{Zhong10}. In Boussinesq studies,
$\text{Ra}^*\sim 1$ is typically found to be a good proxy to separate the
rotation-dominated zonal flow regime (i.e. $\text{Ra}^*
\ll 1$) from the buoyancy-dominated flow regime (i.e. $\text{Ra}^* \gg 1$)
\citep{Gilman77, Aurnou07, Evonuk12}.

In contrast to the $\text{Ra}^*\sim 1$ zonal flow transition in Boussinesq spherical
shells, there is not yet a consensus concerning the mechanisms that
control the breakdown of the smaller-scale convection columns 
\citep[e.g.][]{Schmitz09,King12,King12a,Julien12, Julien12a}.
However, in low Ekman number Cartesian Boussinesq simulations, it is clear 
that
the transition $\text{Ra}^*$ values at which columnar modes become unstable is
significantly less than unity. This low $\text{Ra}^*$ breakdown
criterion suggests that independent behavior transitions may exist for the
large-scale zonal flows and the local-scale columnar convection modes.

In our simulations, however, the convection columns and the zonal
flows seem to undergo simultaneous behavioral transitions. This may occur due to
the moderate Ekman values at which we carry out our suite of simulations, or due
to a difference in the stability properties of convection columns in spherical
shell convection in the presence of zonal flows. Thus, we will focus here only
on the one fundamental transition that separates the rapidly-rotating
regime, in which columns exist and the equatorial zonal flow tends to be
prograde \citep[e.g.][]{Gastine12}, and the buoyancy-dominated regime, in which
the columns are unstable and the equatorial zonal flows tend to be retrograde
\citep[e.g.][]{Aurnou07}.

\subsection{Numerical methods and boundary conditions}

The numerical simulations of this parameter study have been carried out using
the anelastic version of the code MagIC \citep{Wicht02,Gastine12}, which has been
validated in the \cite{Jones11} anelastic dynamo benchmark study.
To solve the system of equations (\ref{eq:anel}-\ref{eq:entropy}), $\rb\vec{u}$
is decomposed into a poloidal and a toroidal contribution

\begin{equation}
 \rb \vec{u} = \vna\times\lp\vna\times W\, \vec{e_r}\rp +
\vna\times Z\, \vec{e_r}.
\label{eq:decomposition}
\end{equation}
$W$, $Z$, $s$ and $p$ are then expanded in spherical harmonic functions up to
degree $\ell_\text{max}$ in colatitude $\theta$ and longitude $\phi$ and in
Chebyshev polynomials up to degree $N_r$ in radius. 
A detailed description of the complete numerical method and the associated
spectral transforms can be found in \cite{Glatz1}. 
Typical numerical resolutions range from ($N_r=65$, $\ell_\text{max}=85$) for
Boussinesq models to ($N_r=161$, $\ell_\text{max}=256$) for the most demanding
anelastic models with $N_\rho=5$. In the latter, a twofold or a fourfold
symmetry in longitude has been used to ease numerical computation. As
rapidly-rotating convection is dominated by high azimuthal wave numbers, this
enforced symmetry is not considered to be influential on the averaged properties
of the flow \citep[e.g.][]{Christensen02, Heimpel05, Jones09}.
Note that this can influence the dynamics at high latitude where
large-scale structures may evolve (e.g. polar vortices). However, this concerns
only a very minor fraction of the total simulated volume.

In all the numerical models presented in this study, we have assumed constant
entropy and stress-free mechanical boundary conditions at both
spherical shell boundaries, $r_i$ and $r_o$.

\subsection{A parameter study}

In this investigation, we consider two different Ekman numbers,
$\text{E}=10^{-3}$ and $\text{E}= 3\times 10^{-4}$, which are
rather moderate but allow us to carry out a significant number
of strongly supercritical cases.  Following our previous
hydrodynamical models, the Prandtl number is set to $1$ and we use a polytropic
index $m=2$ for the reference state.
For all the models of this study, we employ an aspect ratio $\eta=0.6$.
This non-dimensional fluid layer depth exceeds those expected for the
zonal flows in the gas giants
\citep[i.e. $0.85\,R_S$ and $0.96\,R_J$, see][]{Liu08,Heimpel11}. However, this
thicker shell geometry has the advantage of avoiding the significant numerical
expense associated with spectral models of thin shell dynamics.

\begin{table}
\caption{Critical Rayleigh numbers and corresponding azimuthal wave numbers for
the different parameter regimes considered in this study.}
\centering
\begin{tabular}{cccc}
  \toprule
  Ekman number & $N_\rho$ & $\text{Ra}_c$ & $m_{\text{crit}}$ \\
  \midrule
  $10^{-3}$ & 0.01   & $1.412\times 10^{4}$ & 10 \\
  $10^{-3}$ & 1      & $2.919\times 10^{4}$ & 16 \\
  $10^{-3}$ & 3      & $8.126\times 10^{4}$ & 29 \\
  $10^{-3}$ & 5      & $1.624\times 10^{5}$ & 37 \\
 \midrule
  $3\times10^{-4}$ & 0.01 & $5.075\times 10^{4}$ & 15 \\
  $3\times10^{-4}$ & 1    & $1.291\times 10^{5}$ & 24 \\
  $3\times10^{-4}$ & 3    & $3.714\times 10^{5}$ & 47 \\
  $3\times10^{-4}$ & 5    & $6.373\times 10^{5}$ & 60 \\
\bottomrule
 \end{tabular}
\label{tab:rayleigh}
\end{table}

We have performed numerical simulations with various density contrasts spanning
the range from $N_\rho = 10^{-2}$ (i.e. nearly Boussinesq) to $N_\rho=5$ (i.e.
$\rho_{\text{bot}}/\rho_{\text{top}} \simeq 150$).
The strongest stratification considered here is still below the expected density
contrast of the gas giants
interiors: $N_\rho=7.2$
between the 1 bar level and $0.96 R_J$ \citep{Nettelmann12,French12} and
$N_\rho=7.7$ between the 1 bar level and $0.85 R_S$ \citep{Guillot99}.
However, since the density gradient rapidly decreases with depth in both
giant planets, a value of $N_\rho=5$ covers more than $99\%$ of the outward
molecular envelope when starting at $0.96 R_J$ (or $0.85R_S$).
For each density stratification and Ekman number, we vary the Rayleigh number
from onset of convection to $\rasmid \sim 10$. Critical Rayleigh have been
obtained with the linear stability code by \cite{Jones09a} and are given in
Tab.~\ref{tab:rayleigh}.

Most of the runs have been initiated from a conductive thermal state (see
Eq.~\ref{eq:raloc}) with a superimposed random entropy perturbation. For the
most demanding high $\rasmid$ cases, a converged solution at different
parameter values has been used as a starting
condition. Altogether, more than 125 simulations have been performed, each
running for at least $0.3$ viscous diffusion time such that a nonlinearly
saturated state is reached (see Tab.~\ref{tab:results}).

\section{From rotation- to buoyancy-dominated regime}
\label{sec:buo}

\subsection{Convective flows}

\begin{figure*}[t]
  \centering
  \includegraphics[width=8.8cm]{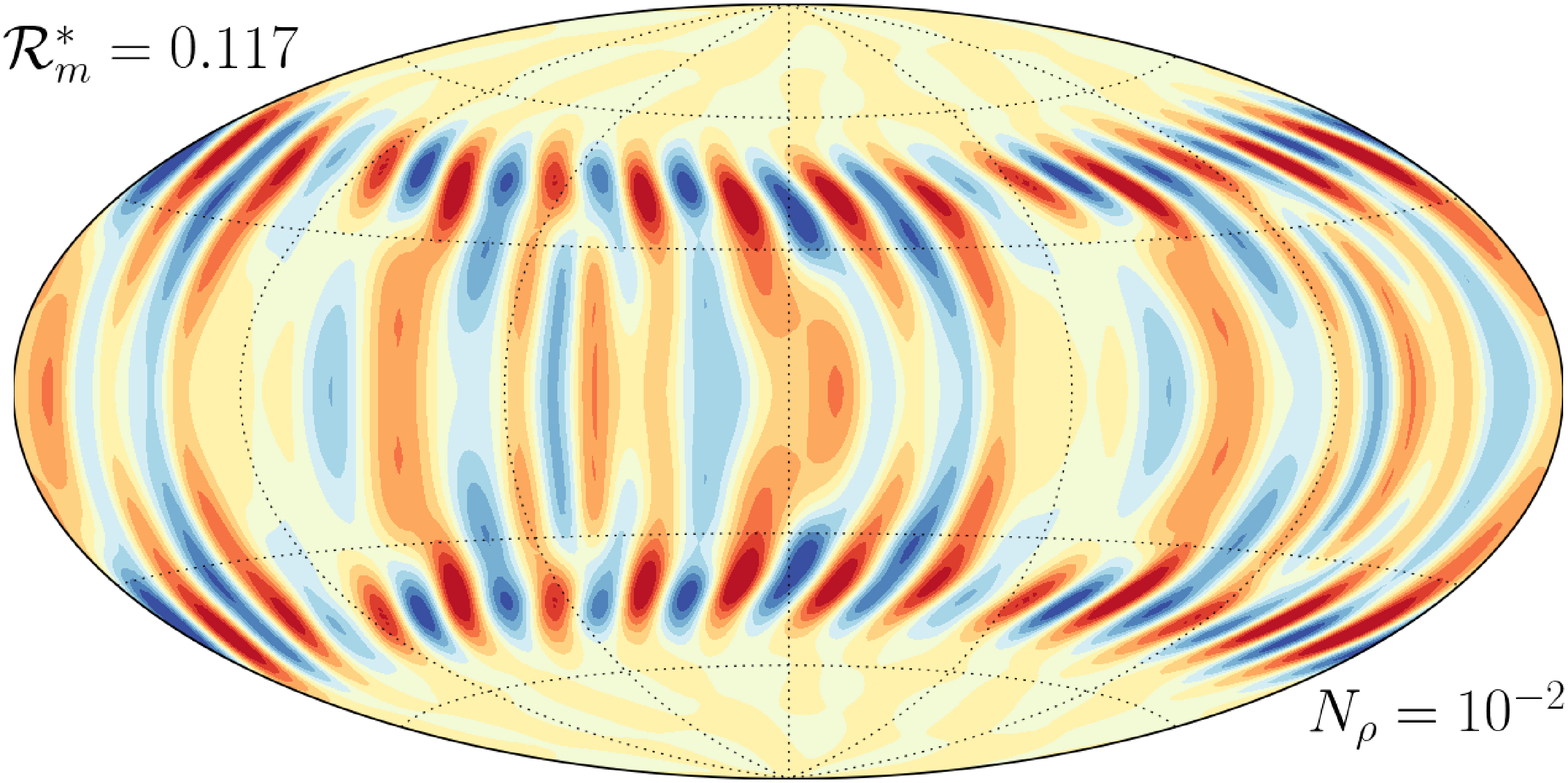}
  \includegraphics[width=8.8cm]{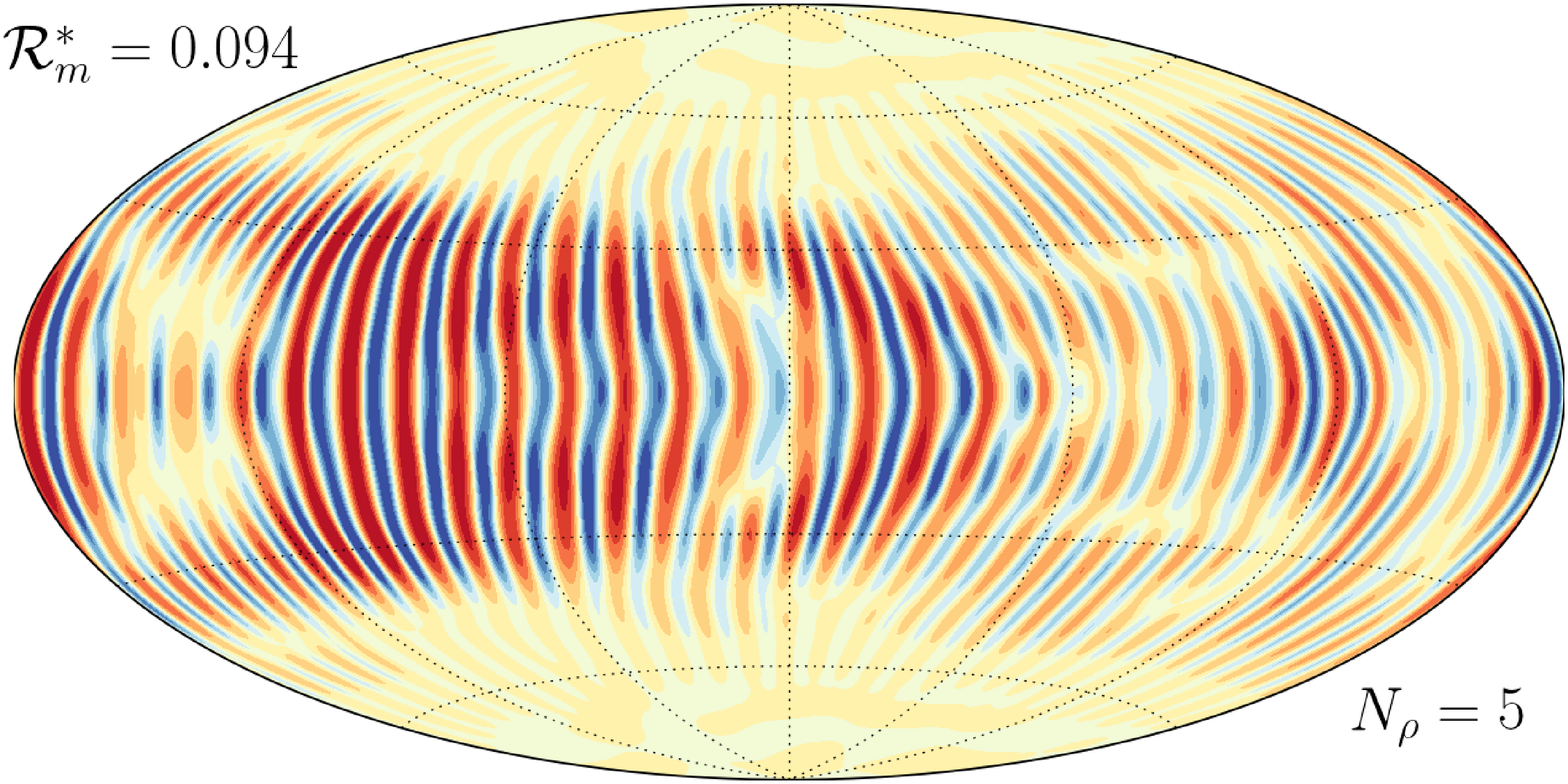}
  \includegraphics[width=8.8cm]{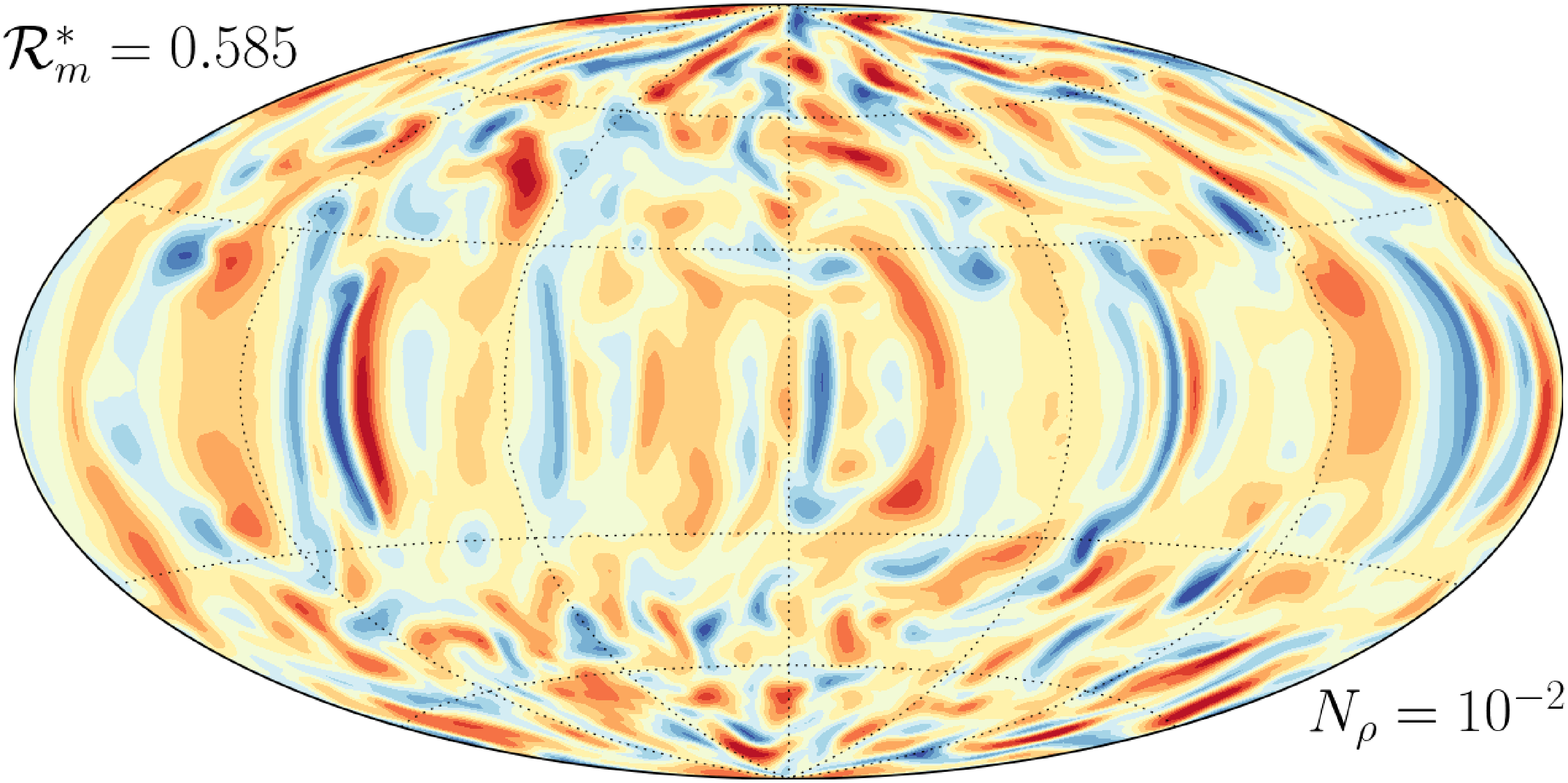}
  \includegraphics[width=8.8cm]{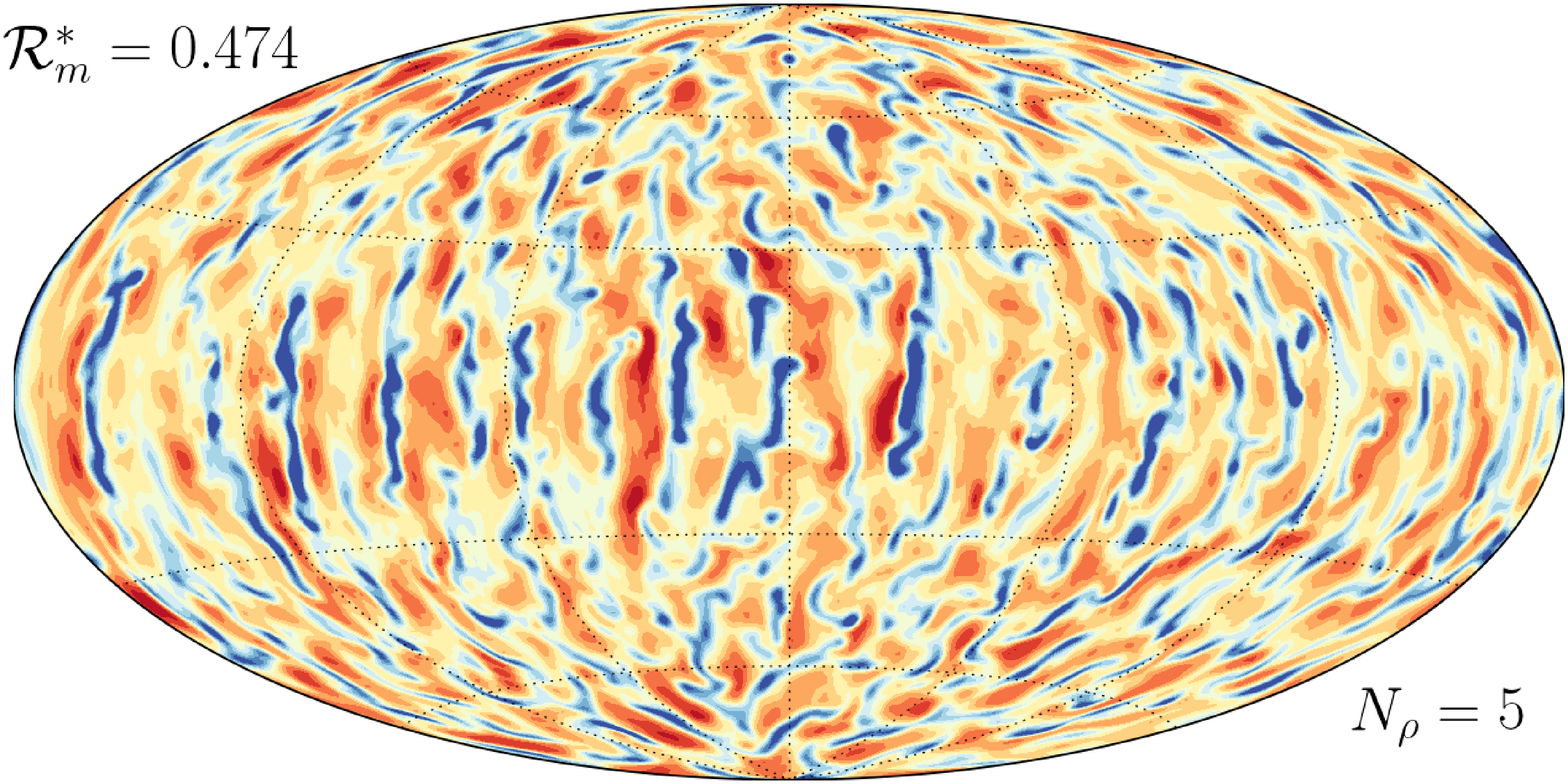}
  \includegraphics[width=8.8cm]{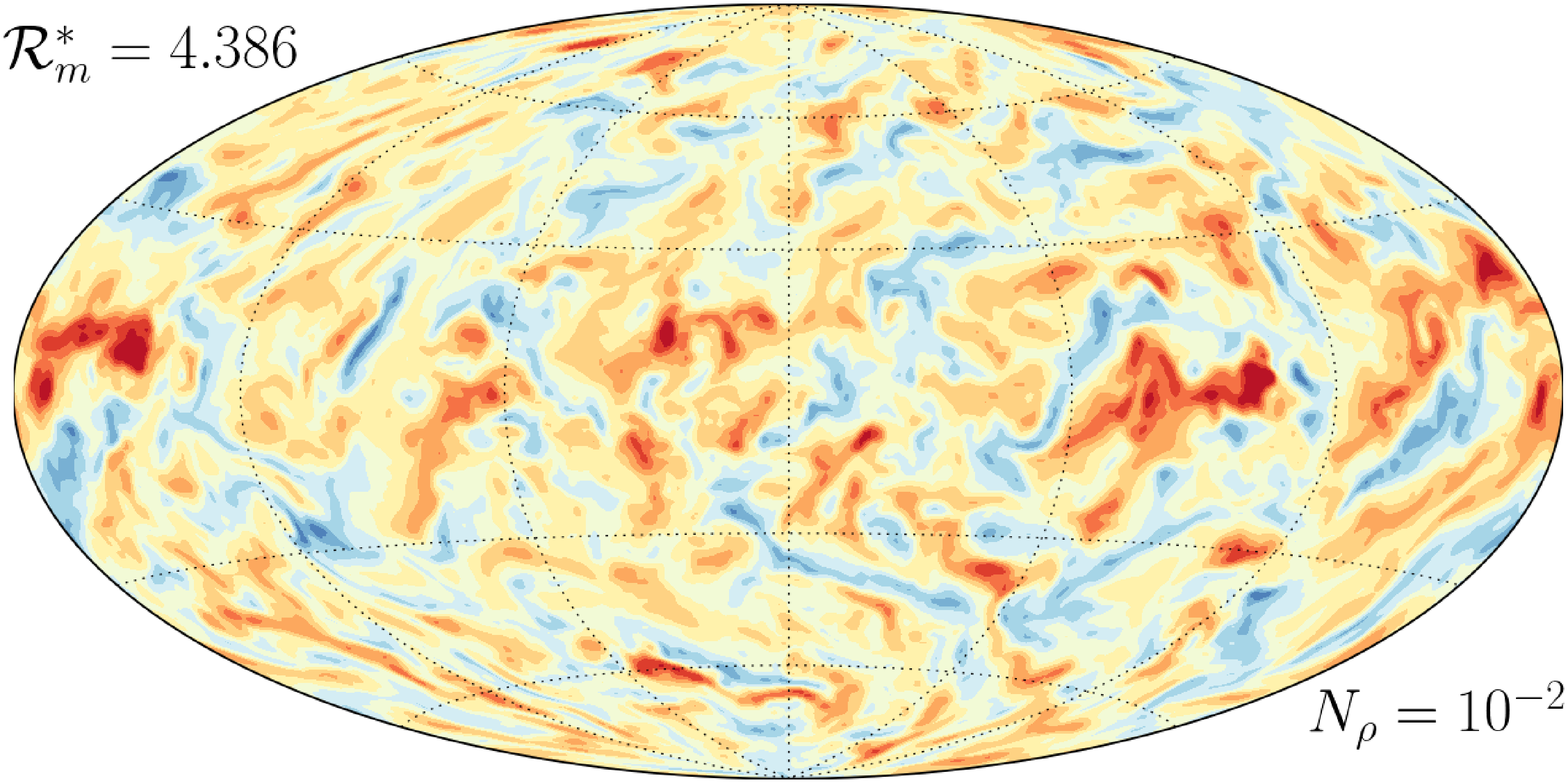}
  \includegraphics[width=8.8cm]{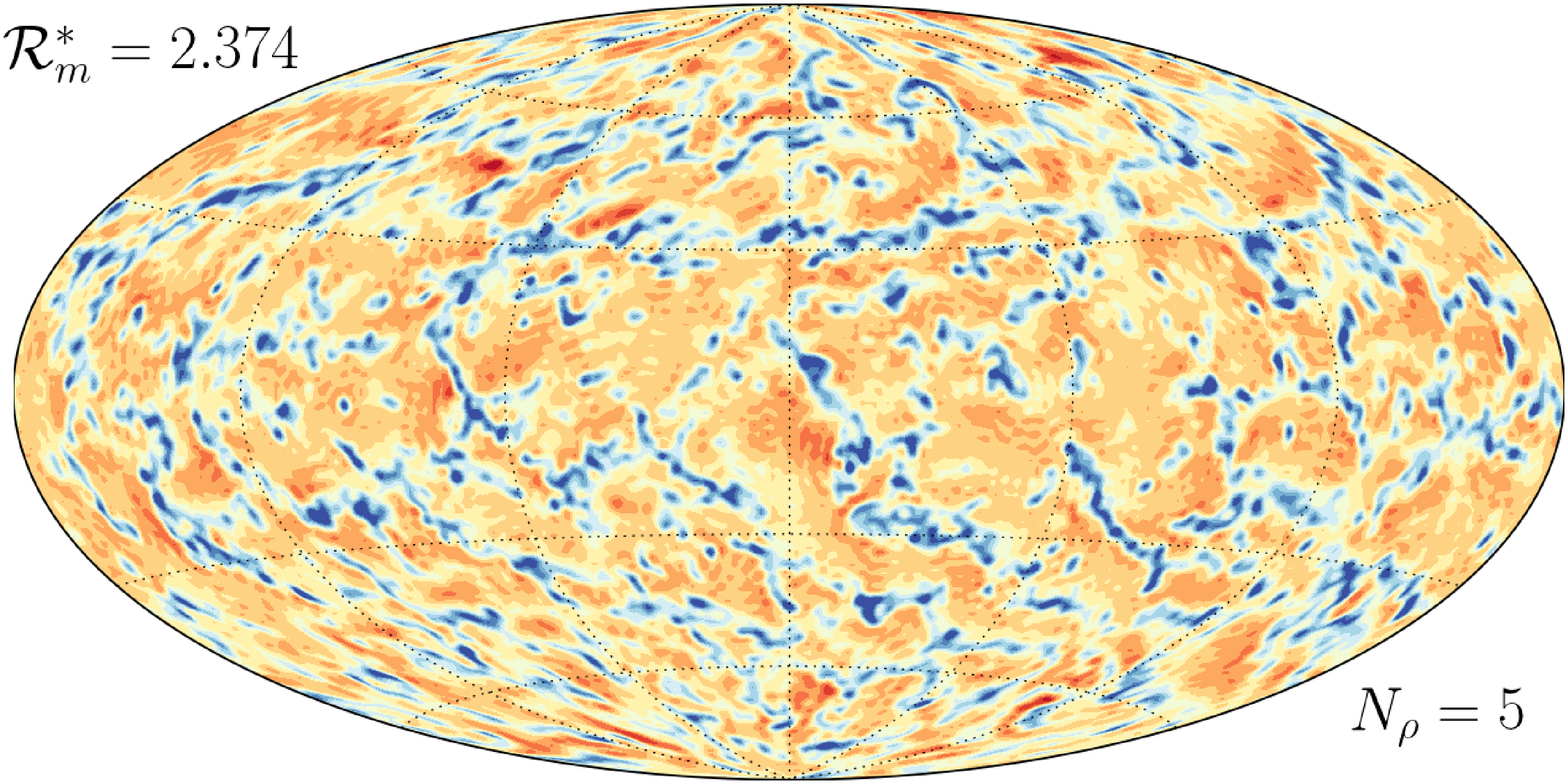}
  \caption{Radial velocity $u_r$ at $r = 0.9\, r_o$
for six different numerical simulations with $\text{E}=10^{-3}$. Boussinesq
models (i.e. with $N_\rho = 10^{-2}$) with  increasing $\rasmid$ are
displayed in the left panels, while anelastic models with $N_\rho=5$ are
displayed in the right panels. Outward flows are  rendered in red, inward
flows in blue.}
  \label{fig:vr}
\end{figure*}

Figure~\ref{fig:vr} shows the radial velocity patterns
when the supercriticality is gradually increased for two different
density stratifications. In the rotation-dominated regime (i.e.
$\rasmid \ll 1$, upper panels of Fig.~\ref{fig:vr}), the convective
structures
are aligned with the rotation axis following the Taylor-Proudman theorem.
The azimuthal lengthscale of the convective
columns is roughly three times smaller in the strongly stratified than
in the Boussinesq model, following the critical wave numbers listed in
Tab.~\ref{tab:rayleigh}. As shown in the linear stability analysis by
\cite{Jones09a}, this variation is due to the background density
contrast that confines convection close to the outer boundary when
$N_\rho$ increases. This confinement causes the decrease in the typical length
scales in both the radial and the azimuthal directions \citep[see
also][]{Gastine12}.

Convection develops both inside and outside the tangent
cylinder in cases with more supercritical Rayleigh numbers that still have
$\rasmid \lesssim 1$ (Fig.~\ref{fig:vr}, middle row). The integrity of the
convective columns is disturbed due to the gradual loss of geostrophy as the
buoyancy forcing increases in strength \citep[e.g.][]{Soderlund12}.
The difference in lengthscales between the two cases is roughly retained
at $r=0.9\, r_o$. However, due to the local variations of the
density scale height, the convective flow lengthscale increases in the deep
interior of the $N_\rho=5$ case \citep[see for instance Fig.~5
in][]{Gastine12}.

The alignment of the convective features along the rotation
axis is completely lost when buoyancy dominates the global-scale force balance
($\rasmid \gtrsim 1$, lower panels of Fig.~\ref{fig:vr}). While up- and
downwellings have a very similar structure
in the Boussinesq model, a strong asymmetry is visible in the
strongly stratified case. Here the convection roughly forms a network
of thin elongated downflows that enclose broader and weaker upflows. Upwelling
structures tend to expand and acquire a mushroom-like shape, while downwelling
plumes are narrow and concentrated. This network-like pattern of convection has
been frequently observed in numerical models of the solar
granulation \citep[e.g.][]{DeRosa02,Miesch08}.

\subsection{Zonal winds}

\begin{figure*}[t]
 \centering
  \includegraphics[width=17cm]{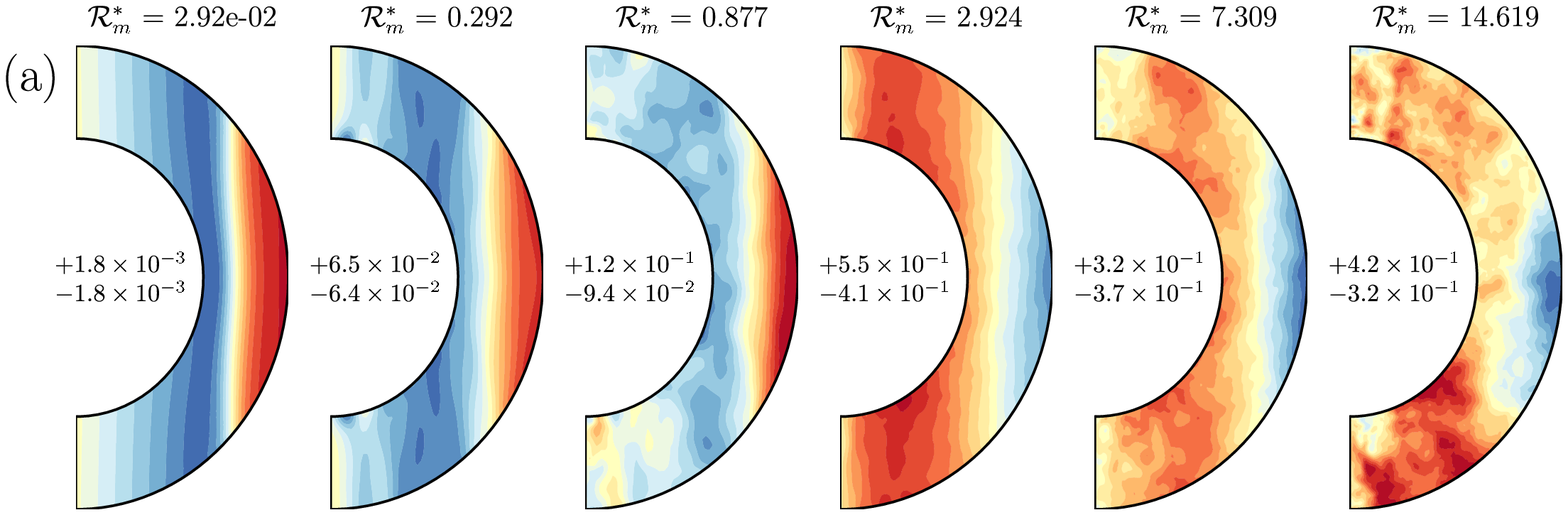}
  \includegraphics[width=17cm]{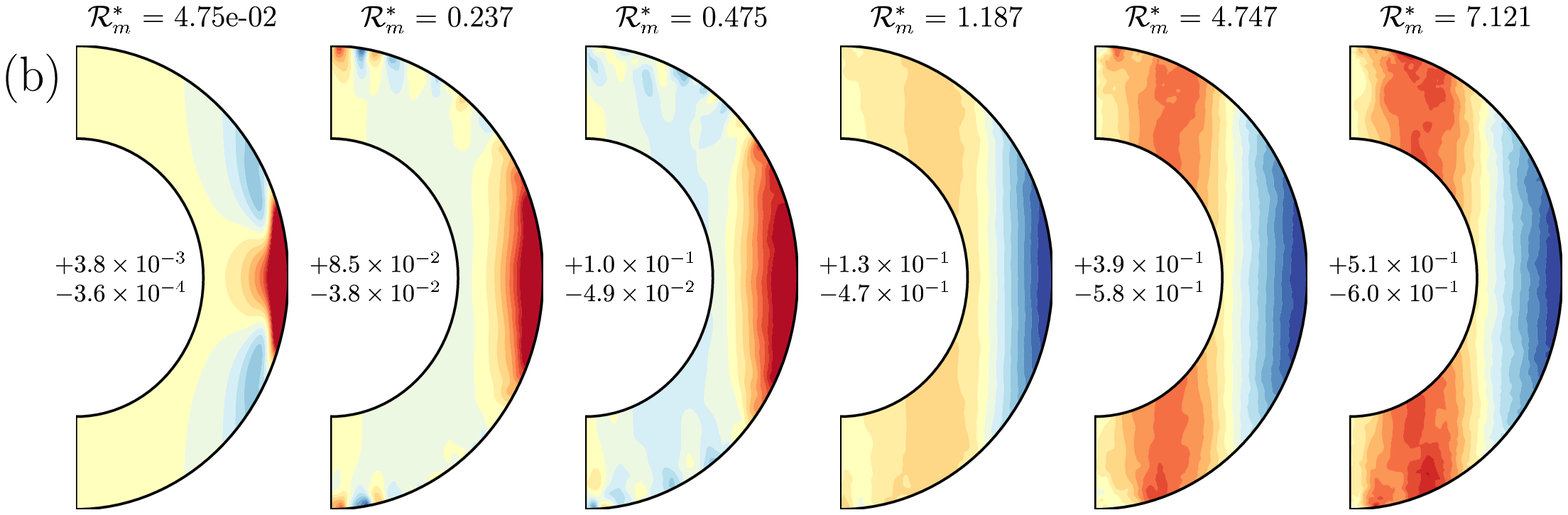}
  \caption{Zonally averaged azimuthal velocity in the meridian plane for 
simulations with $\text{E}=10^{-3}$ and increasing 
$\rasmid$. Upper panels correspond to Boussinesq models (i.e. $N_\rho =
10^{-2}$) and lower panels to $N_\rho=5$.
Colorscales are centered around zero: prograde jets are 
rendered in red, retrograde jets in blue. In some cases, the prograde 
contours have been truncated in amplitude
to emphasise the structure of the retrograde flows. Extrema of the zonal flow 
velocity are indicated in the center of each panel (velocities are expressed 
in Rossby number units, i.e. $u/\Omega r_o$).}
 \label{fig:vp}
\end{figure*}

Figure~\ref{fig:vp} shows the change in the zonal winds when
$\rasmid$ is increased in the nearly Boussinesq models ($N_\rho=10^{-2}$, left
panels) and in the strongly stratified models ($N_\rho=5$, right panels).
The typical pattern of an eastward (i.e., prograde) outer and a westward inner
geostrophic zonal flow already develops at mildly supercritical Rayleigh numbers
(top row). These zonal flows are driven by Reynolds stresses (i.e. the
statistical correlation of convective flow components), which rely on the
prograde tilt of the convective columns induced by the boundary curvature
\citep[e.g.][]{Busse83,Zhang92, Christensen02}. In the rotationally-dominated
models, Reynolds stresses generate a positive
angular momentum flux away from the rotation axis that is balanced by viscous
drag. Because of the strong confinement of convection close to the outer
boundary in the $N_\rho=5$ case, the equatorial jet is narrower than in the
Boussinesq model. In addition, its amplitude is significantly larger than the
amplitude of the adjacent retrograde jet in the stratified model. 

With further increasing supercriticality, the differences discussed above 
tend
to vanish. Thus, for $\rasmid \lesssim 0.5-1$ (third panels), the equatorial
jets have indeed very similar amplitudes and latitudinal extent while the
retrograde jet is still somewhat weaker in the anelastic model.
This agrees with the results or our previous parameter study in 
the $\rasmid \lesssim 1$ regime \citep{Gastine12}.
Note that the larger Ekman number cases considered here do
not allow for the multiple jets solution found in \cite{Gastine12},
where the main equatorial jet was flanked by a retrograde zonal flow attached to
the tangent cylinder and a pair of high latitude prograde jets.

When buoyancy starts to dominate the force balance (i.e. $\rasmid \gtrsim
1$), the zonal flow direction largely reverses. The flow outside the tangent
cylinder becomes mainly retrograde and the flow inside the tangent cylinder
prograde. The amplitude of these zonal winds also increases
roughly by a factor $5$ at this transition reaching $\text{Ro} \simeq -0.5$.
The zonal winds are nearly $z$-independent for $\rasmid \simeq 1$ (fourth
panels), even though the convective motions are no longer in geostrophic
balance. However, a further increase of the supercriticality produces some
pronounced smaller-scale $z$-dependent features in the Boussinesq models. These
small-scale structures are highly time-dependent and would therefore cancel out
when time-averaged quantities are considered. The last panels of
Fig~\ref{fig:vp} show that both the mean zonal flow and the small-scale
ageostrophic motions depend on the density contrast in the $\rasmid \gg 1$
regime.

\begin{figure}
   \centering
   \includegraphics[width=8.8cm]{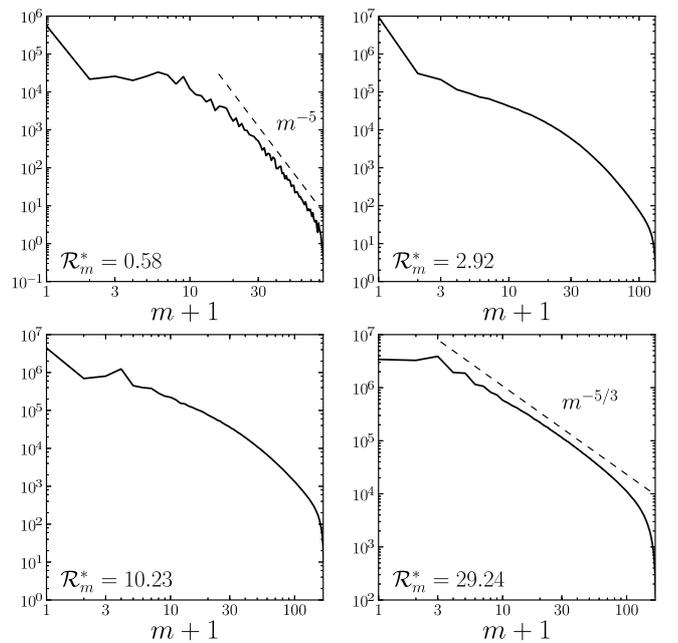}
   \caption{Time-averaged toroidal kinetic energy spectra for four
   different numerical models with $N_\rho=10^{-2}$ and $\text{E}=10^{-3}$.}
   \label{fig:spec}
 \end{figure}

This change in the zonal flow structure is confirmed by the
time-averaged toroidal kinetic energy spectra shown in Fig.~\ref{fig:spec} for
four Boussinesq models at $\text{E}=10^{-3}$. In the rotation-dominated regime
($\rasmid < 1$, upper left panel), the spectrum peaks at $m=0$,
illustrating the strong contribution of axisymmetric zonal flows in the
toroidal kinetic energy budget. For spherical harmonic orders $m > 20$, the
spectra follow a clear power-law behaviour and the steep decrease (close to
$m^{-5}$) is very similar to previous quasi-geostrophic studies computed
at much lower Ekman numbers ($\text{E} = 10^{-8}$)
\citep{Schaeffer06} and matches the $k^{-5}$ scaling derived by \cite{Rhines75}
in the $\beta$-plane turbulence framework. Axisymmetric zonal flows still
dominate the toroidal energy in the $\rasmid =
2.92$ and $\rasmid = 10.23$ cases, but the slope at
smaller lengthscales flattens gradually. At $\rasmid=29.24$ (lower right panel),
the spectrum shows a mild maximum at $m=2$ before clearly following an inertial
range scaling of $m^{-5/3}$, the theoretical behaviour expected for homogeneous
and isotropic 3-D turbulence \citep[e.g.][]{Lesieur}.
In this case, the axisymmetric zonal flow contribution is not dominant anymore
as already suggested by the last panel of Fig.~\ref{fig:vp}a.

\subsection{Convection regimes}

\begin{figure}[t]
 \centering
  \includegraphics[width=8.5cm]{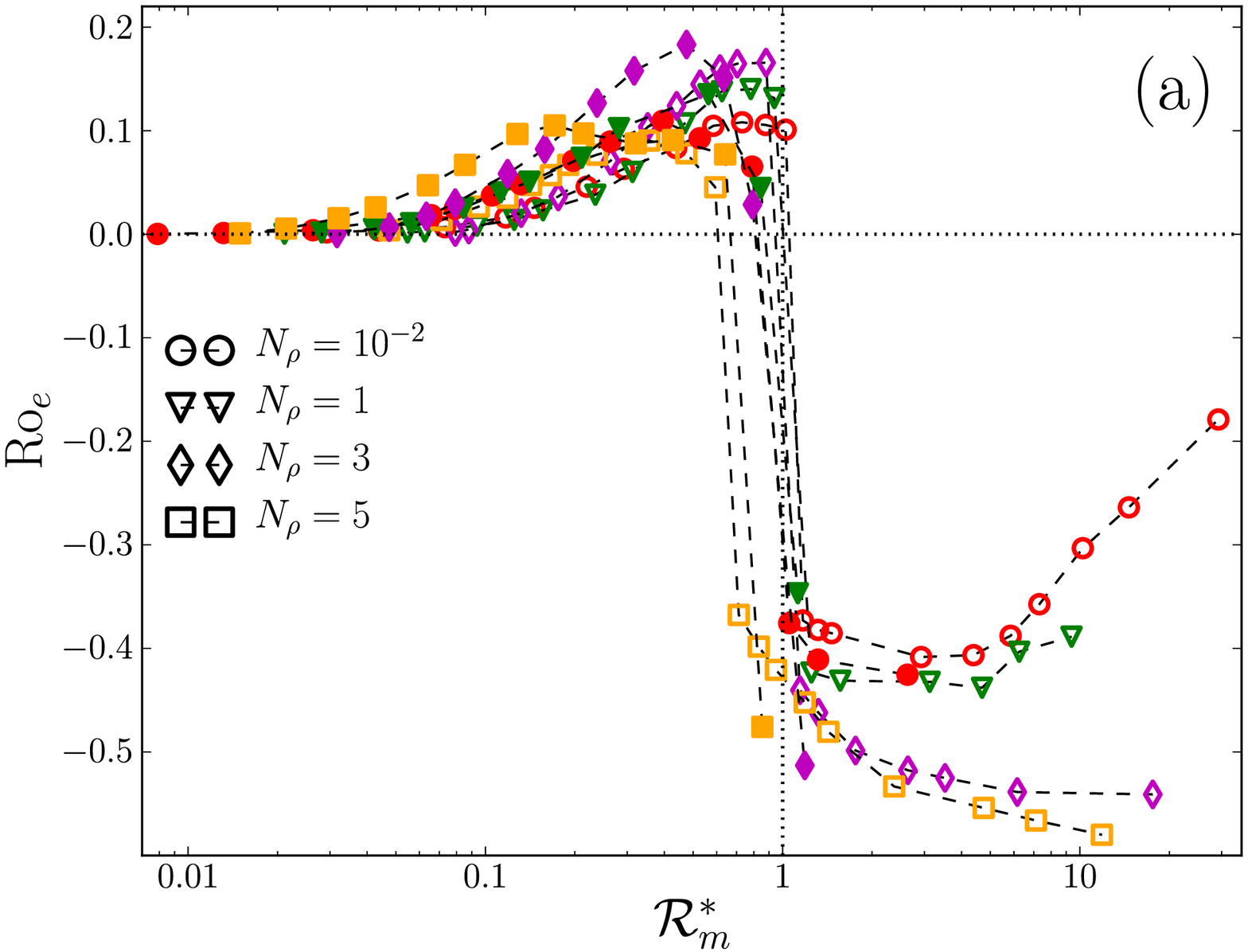}
   \includegraphics[width=8.5cm]{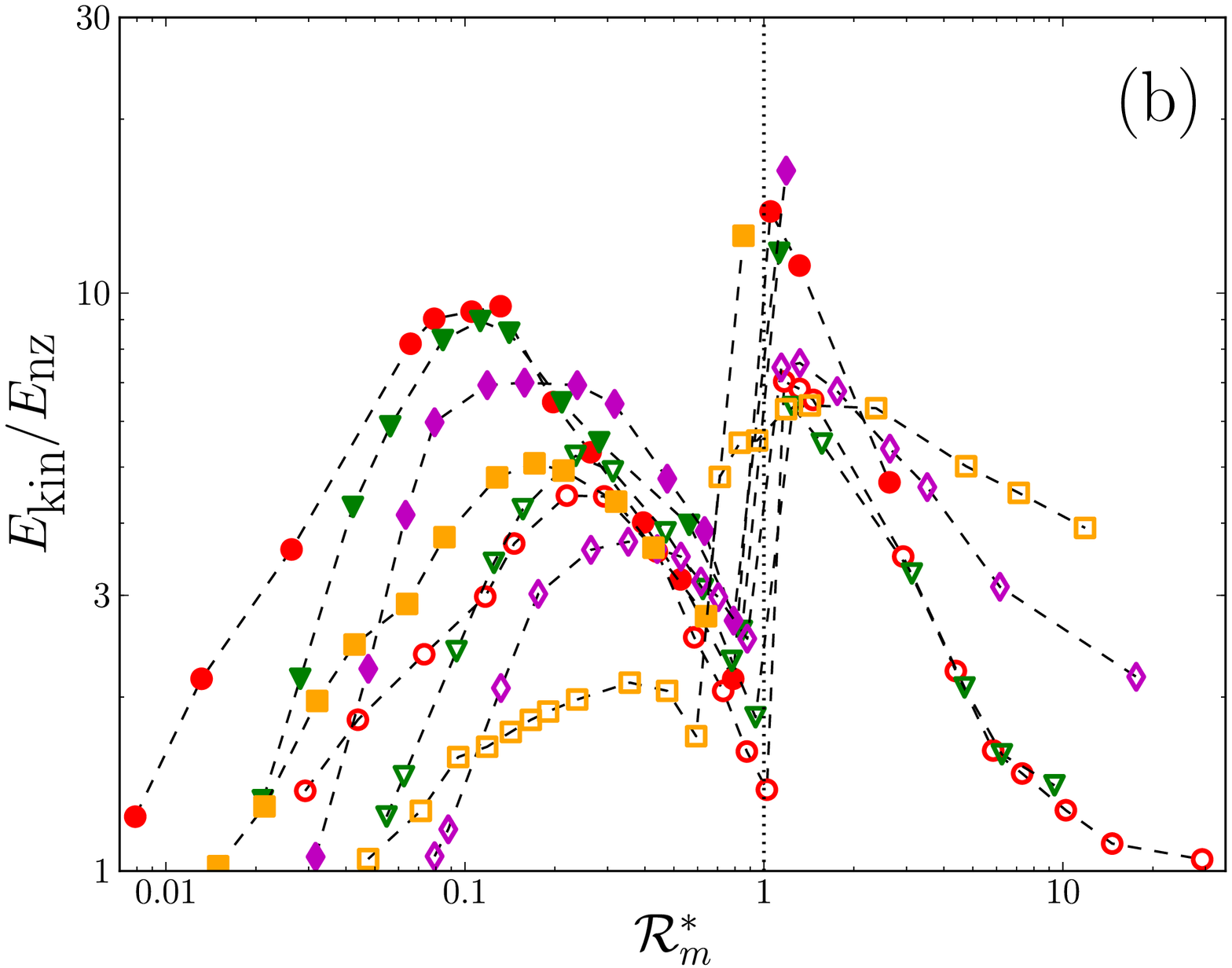}
  \caption{(a) Amplitude of the surface zonal wind at the
equator (in units of $\text{Ro}_e=u_\phi/\Omega r_o$) plotted against
$\rasmid$.
(b) Ratio of total over non-zonal kinetic energy plotted against
$\rasmid$. In both panels, closed symbols correspond to simulations with
$\text{E}=3\times 10^{-4}$ and open symbols to simulations with
$\text{E}=10^{-3}$. The transition at $\rasmid = 1$ is emphasised by a
dotted vertical line.}
  \label{fig:rossby}
\end{figure}

To further investigate the change in zonal flow regime observed in
Figs.~\ref{fig:vr}-\ref{fig:vp}, we consider time-averaged quantities for the
whole set of numerical models computed in this parameter study. 
We focus in the following on the zonal flow amplitude
characterised by the dimensionless Rossby number
$\text{Ro}=\overline{u}_\phi/\Omega r_o$
and on the contribution of axisymmetric toroidal energy in the total kinetic
energy budget. Following \cite{Christensen02}, this is quantified by the
ratio of the total to the non-zonal kinetic energy $E_\text{kin}/E_\text{nz}$,
where $E_\text{kin}$ has been obtained by

\begin{equation}
 E_\text{kin} = \dfrac{1}{\tau}\dfrac{1}{V}\int_{t_0}^{t_0+\tau}\int_V \rb
\vec{u}^2 \, dV\,dt,
\end{equation}
where $V$ is the volume of the spherical shell and $\tau$ is the time-averaging
interval. Each numerical simulation has been averaged long enough to suppress
the short term variations.

\begin{figure}
 \centering
  \includegraphics[width=8.8cm]{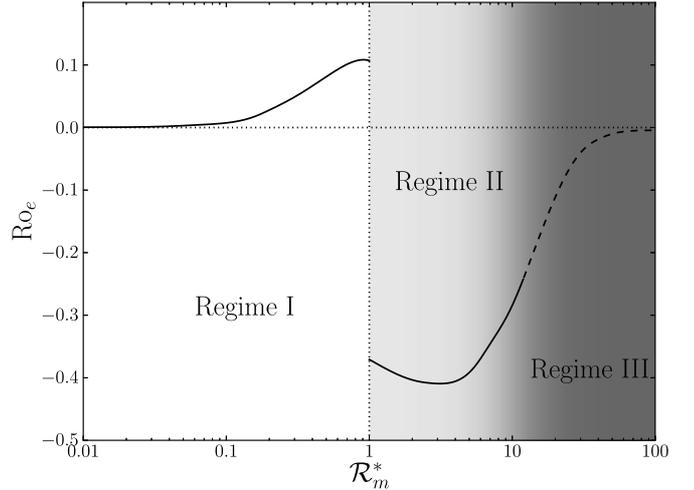}
   \caption{Regime diagram of the mean surface zonal
flow at the equator as a function of $\rasmid$. Regime I corresponds to the
rotation-dominated regime, in which $\rasmid \lesssim 1$. Regime II corresponds
to
the buoyancy-dominated regime in which the mixing of the angular momentum per
unit of mass is efficient, while Regime III suggests a possible suppression of
the zonal flow for $\rasmid \gg 1$. This regime diagram is based on the 
Boussinesq cases from Fig.~\ref{fig:rossby}.}
  \label{fig:regime}
\end{figure}

Figure~\ref{fig:rossby} shows how the surface equatorial zonal flow amplitude
$\text{Ro}_e$ and the ratio
$E_\text{kin}/E_\text{nz}$ change when $\rasmid$ is increased for
numerical models with different Ekman numbers and density contrasts. In the
rotation-dominated regime ($\rasmid \lesssim 1$), the equatorial jets is always
prograde and its amplitude gradually increases with $\rasmid$. A maximum
amplitude of $\text{Ro}_e\simeq 0.1-0.15$ is reached around $\rasmid \simeq
0.5-0.8$. In this regime, the influence of the density contrast is not 
obvious: the $N_\rho=3$ cases produce stronger jets than
the Boussinesq models, which are, rather similar to the
strongly stratified $N_\rho=5$ cases.

Figure~\ref{fig:rossby}b shows that the
ratio $E_\text{kin}/E_\text{nz}$ first increases for weak to moderate
supercriticalities before decaying once convection is strongly driven. This
decay is attributed to the gradual decorrelation of the convective flow
components in the strongly-driven regime, which reduces the efficiency of the
energy transfer between small-scale convection and large-scale mean zonal flows
via Reynolds stresses \citep{Christensen02,Gastine12}. The maximum of this
ratio depends on both the density stratification and the Ekman number. In
Boussinesq models at $\text{E}=3\times 10^{-4}$, this maximum is around 10
(which means to 90\% of  the total kinetic energy is contained in the zonal
flows) while it reaches only 5 (i.e. 80\%) in the $N_\rho=5$ models. In
addition, decreasing the Ekman number tends to produce a stronger zonal flow
contribution \citep[see][]{Christensen02}.

An abrupt transition to retrograde equatorial zonal winds takes place close
to $\rasmid \simeq 1$, independently of the density stratification and the
Ekman number considered. The retrograde equatorial jet amplitude
is roughly multiplied by a factor 4 at this transition to reach $\text{Ro}_e
\simeq -0.4$. This goes along with a larger $E_\text{kin}/E_\text{nz}$ value
that reaches approximately 7 for the $\text{E}=10^{-3}$ cases (i.e. 85\%) and 15
for the $\text{E}=3\times 10^{-4}$ cases (i.e. 93\%).

As $\rasmid$ is increased further, the density stratification starts to
play a more important role.
For stronger stratifications, the equatorial zonal flow amplitude further
increases while it roughly levels off and finally decreases in the Boussinesq
models. This is also reflected in the variation of
$E_\text{kin}/E_\text{nz}$
where the strongly stratified cases decay slower at large $\rasmid$ than the
weakly stratified models. Since we expect that the convective fluctuations will
dominate the mean zonal flows at $\rasmid \gg 1$, the ratio $E_\text{kin}
\simeq E_\text{nz}$ should ultimately tend to unity.
This is already observed in Boussinesq cases in Fig.~\ref{fig:rossby}b where a
value of 1.04 is reached at $\rasmid=29.24$. For the stronger
stratified cases, much larger $\rasmid$ values are required than we could
afford to simulate numerically.

Figure~\ref{fig:regime} shows a tentative regime diagram based on our simulation
results. Regime I corresponds to rotation-dominated cases ($\rasmid<
1$) where convection shows a columnar structure and maintains a prograde
equatorial zonal flow. Regime II is characterised by three-dimensional
convection and a strong retrograde equatorial jet. The transition between
regimes I and II takes place around $\rasmid\sim 1$ independently of the
density stratification and the Ekman number.
The strong decrease of the zonal flow amplitude observed in the Boussinesq
models for $\rasmid > 5$ points toward a possible third regime in which the
zonal flows are insignificant and convection becomes isotropic. The
transition between regimes II and III is however gradual and rather difficult
to pinpoint (emphasised by the dashed-line and the color gradient in
Fig.~\ref{fig:regime}). Our simulations suggest for instance that the transition
depends on the density stratification.

The reason for the density-dependent jets amplitude observed in the
buoyancy-dominated regime and the possibility of a third regime
at large $\text{Ra}^*$ are further discussed in the next section.

\section{Angular momentum mixing}
\label{sec:amm}

\subsection{Influence of the density stratification}

\begin{figure}
 \centering
 \includegraphics[width=8.5cm]{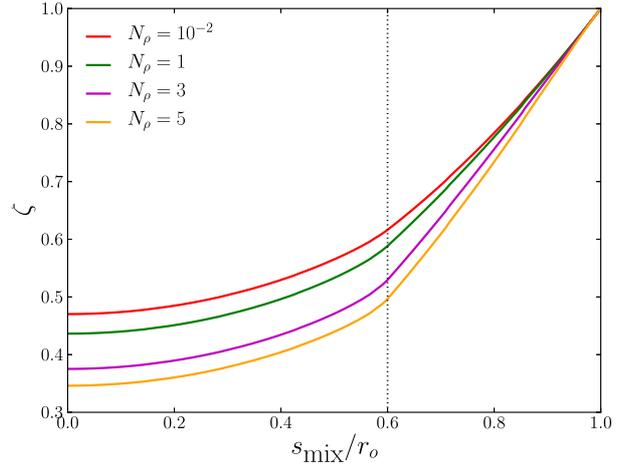}
 \caption{Profiles of $\zeta$ for different density
stratifications as a function of $s_\text{mix}$, the minimum cylindrical radius
above which mixing of angular momentum per unit of mass occurs. The vertical
line corresponds to the location of the tangent cylinder.}
 \label{fig:mstarNrho}
\end{figure}

When buoyancy dominates Coriolis force, strong convection can locally mix the
angular momentum. Flows in which angular momentum is indeed homogenised have
been found in several studies of atmospheric dynamics
or numerical models of rotating convection with a weak rotational influence
\citep[e.g.][]{Gilman77,Hathaway82,DeRosa02,Aurnou07,Bessolaz11}.

\begin{figure*}[t!]
  \centering
  \includegraphics[width=8cm]{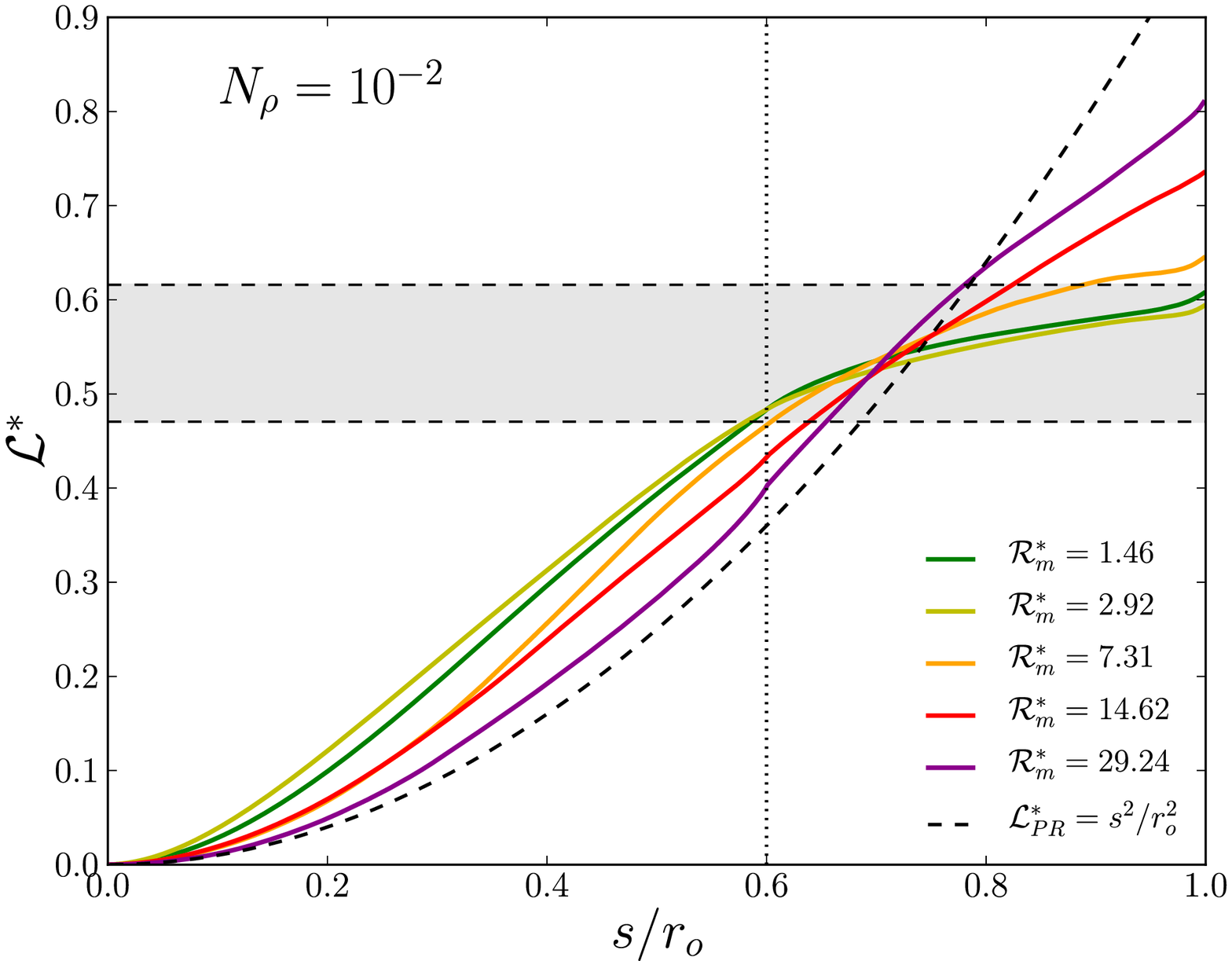}
  \includegraphics[width=8cm]{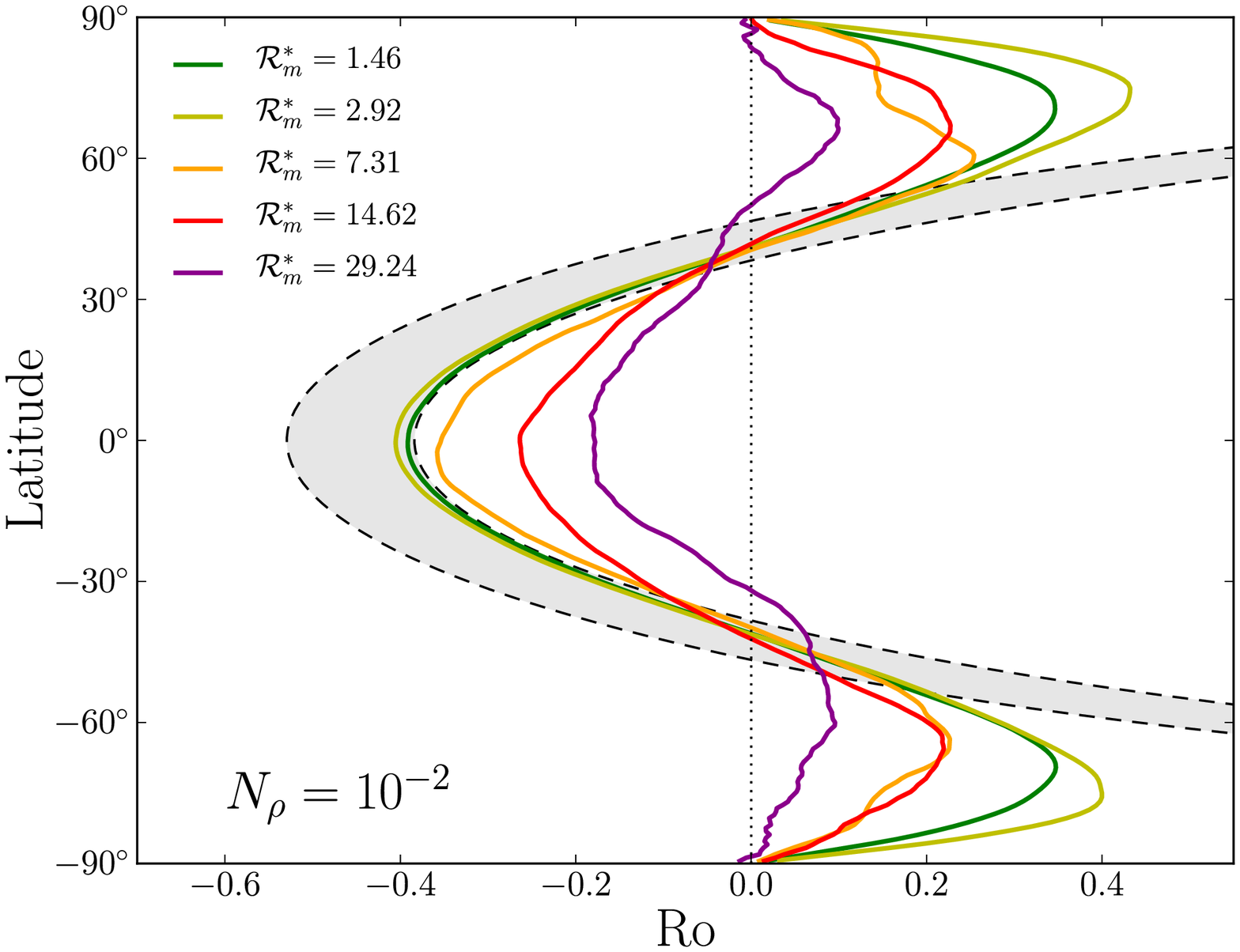}
  \includegraphics[width=8cm]{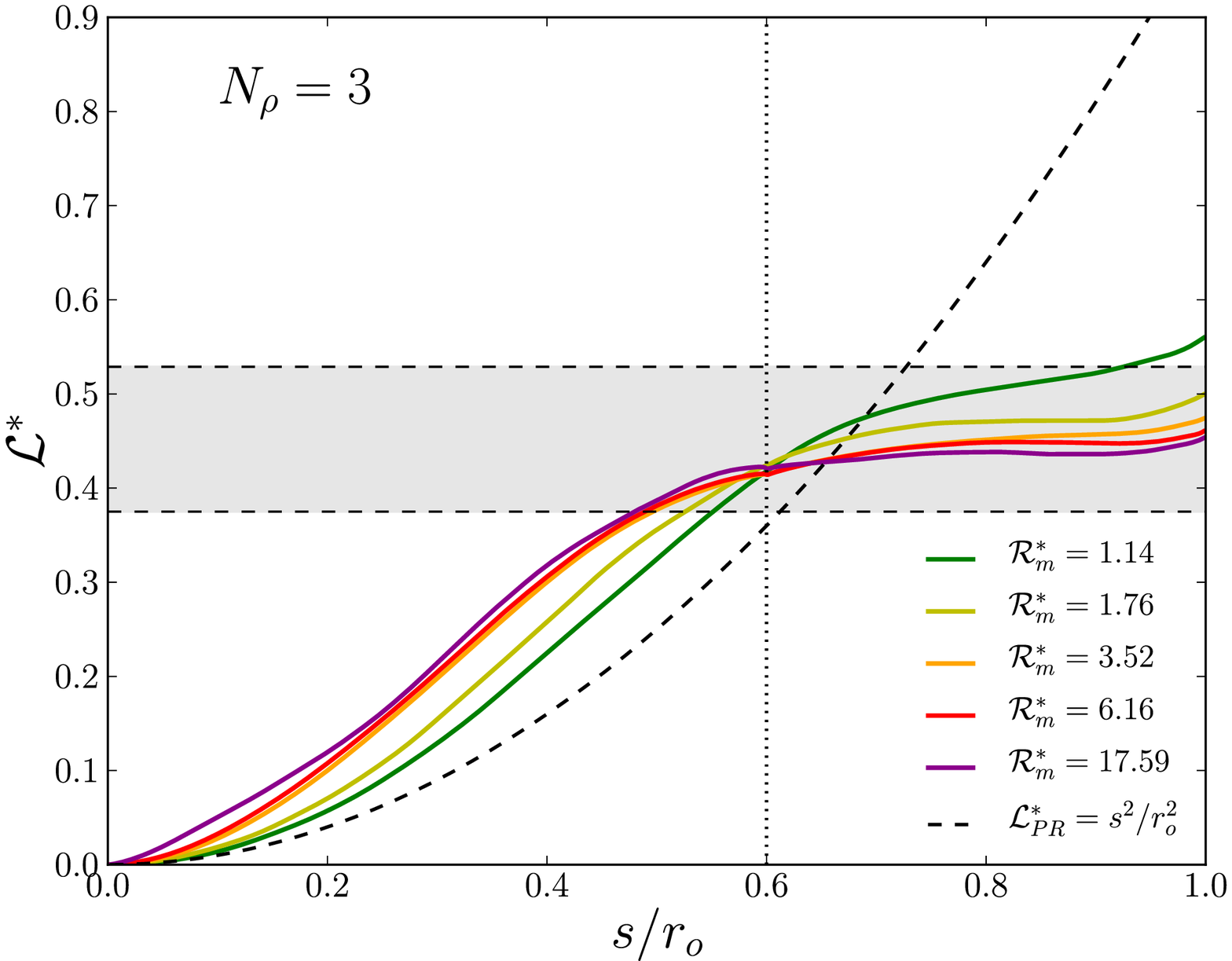}
  \includegraphics[width=8cm]{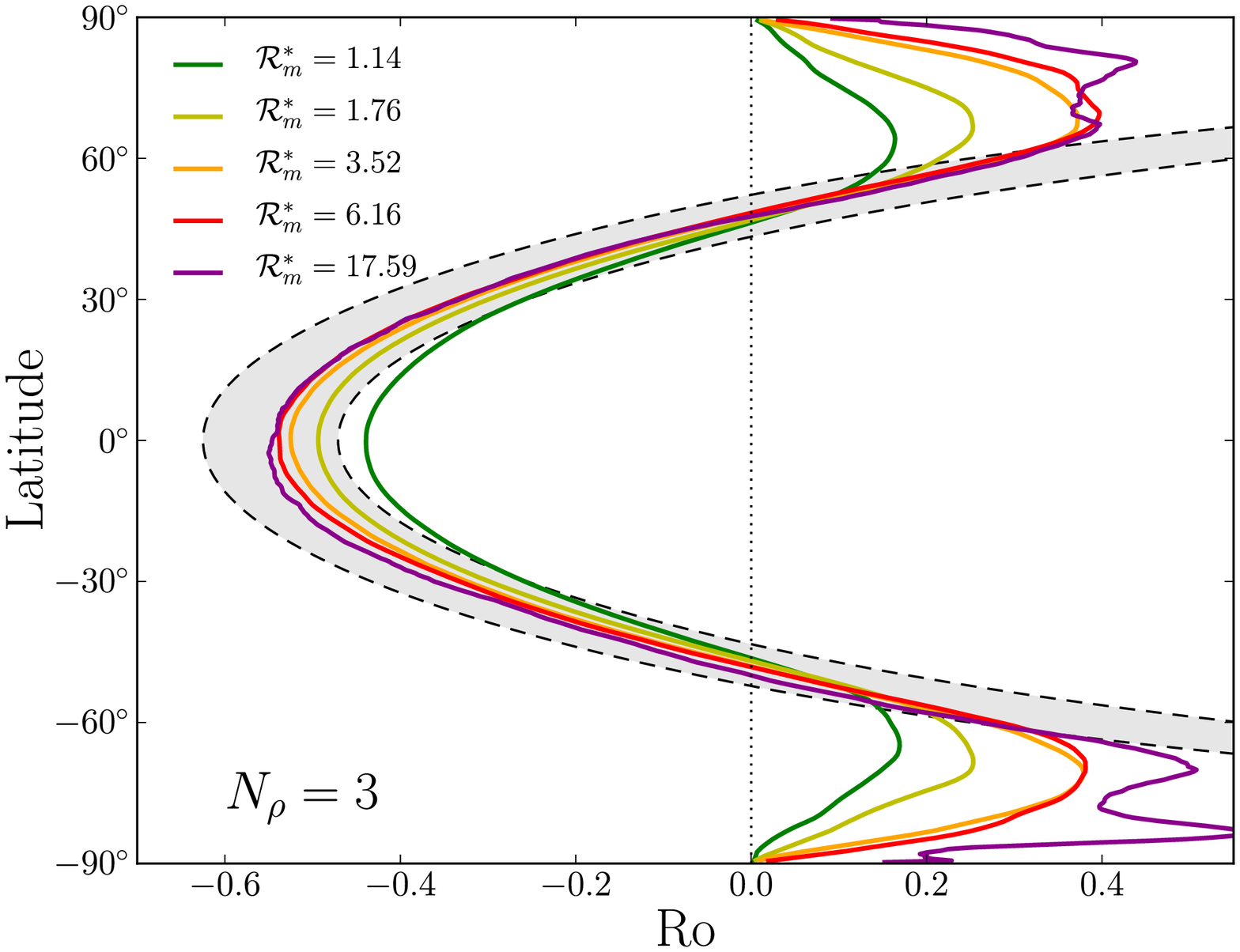}
  \includegraphics[width=8cm]{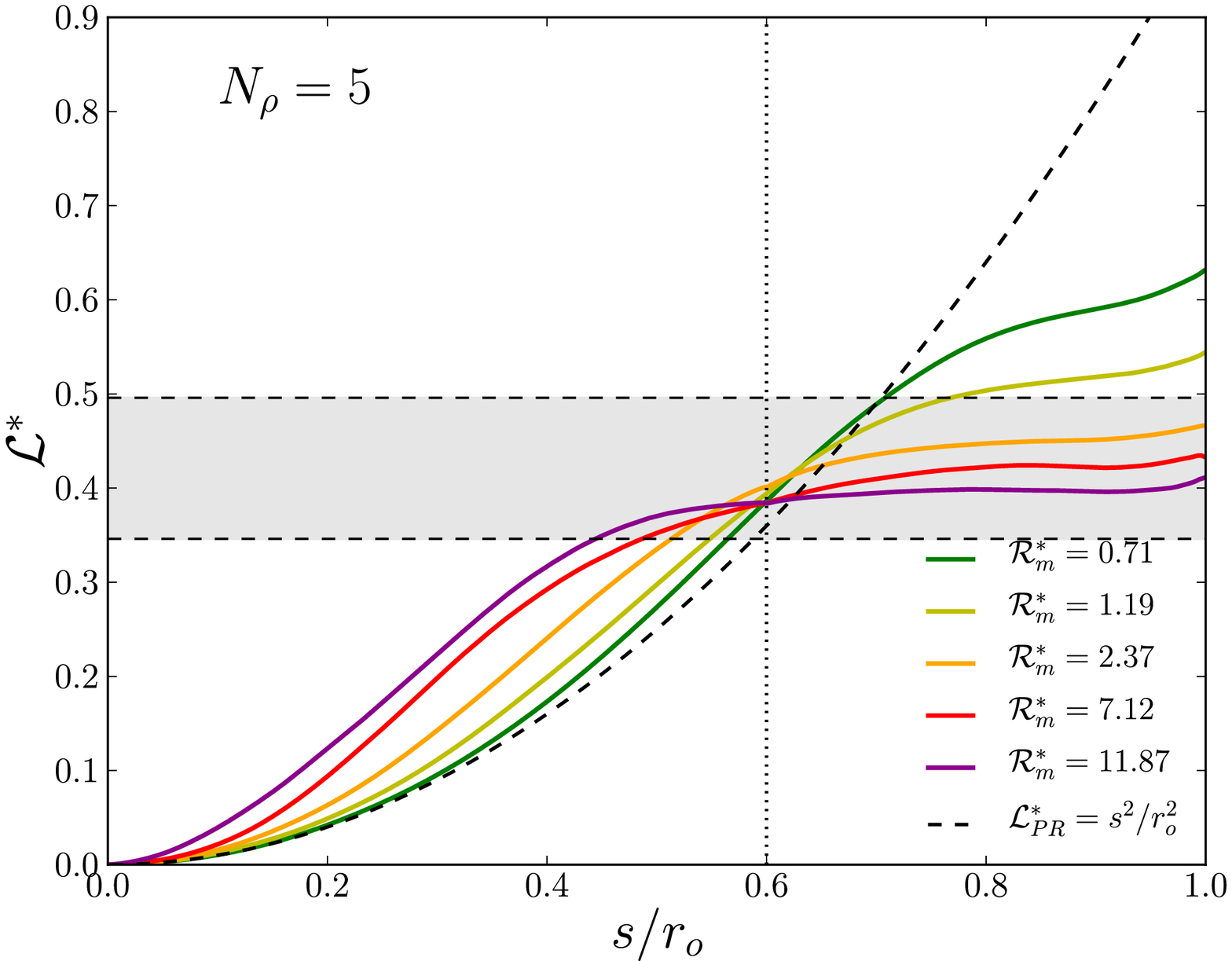}
  \includegraphics[width=8cm]{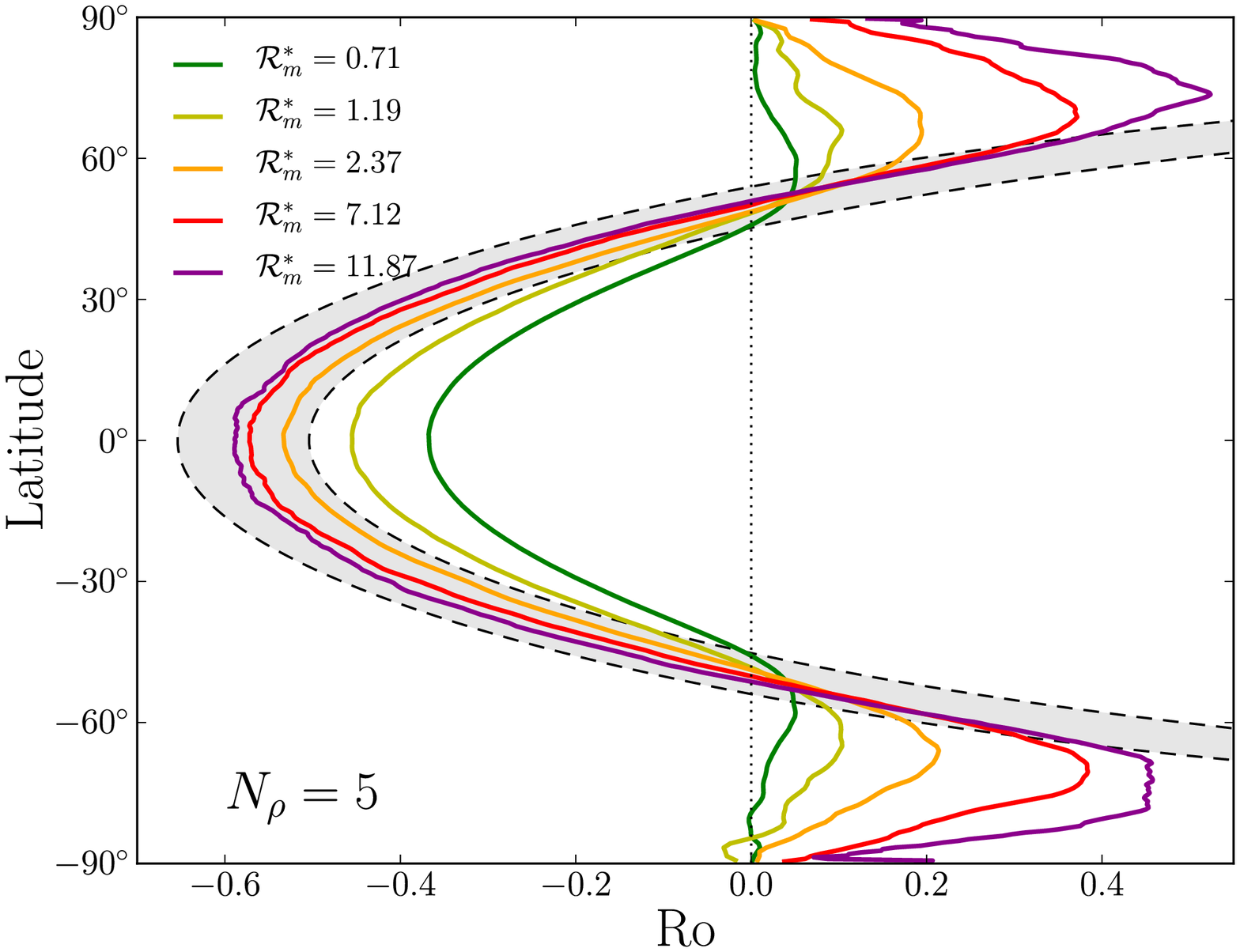}
  \caption{ Left: time-averaged specific angular momentum
plotted against the cylindrical radius $s$ for various numerical simulations
with $\text{E}=10^{-3}$. Right: corresponding azimuthally-averaged velocity
profiles as a function of latitude at the outer boundary. The vertical lines in
the left panels correspond to the tangent cylinder. For comparison, the
grey-shaded area are delimited by the theoretical values associated with ${\cal
L}^*$ mixing over the whole spherical shell (lower bound) and outside the
tangent cylinder only (upper bound, see Tab.~\ref{tab:mstar} for the
corresponding values).}
  \label{fig:mom}
\end{figure*}

In an inertial reference frame, the angular momentum of a fluid parcel is given
by

\begin{equation}
 {\cal M} = {\cal M}_\text{ZF}+{\cal M}_\text{PR}=\rho \overline{u}_\phi
s\,\delta{\cal V} +
\rho \Omega s^2\,\delta{\cal V},
\end{equation}
where $\delta{\cal V}$ is the volume of the fluid element, $s=r\sin\theta$ is
the cylindrical radius, ${\cal M}_\text{ZF}$ is the
angular momentum due to zonal flows and ${\cal M}_\text{PR}$ is the angular
momentum that comes from the planetary rotation. The overbars here
indicate an azimuthal average. 
Because of the anelastic continuity equation~(\ref{eq:anel}), the mass of a
fluid element is conserved during its displacement such that
 $\rho \delta{\cal V}=\text{const.}$. If, in addition, a fluid parcel conserves
its angular momentum ${\cal M}$, then, as hypothesized
by \cite{Gilman79}, the angular momentum \emph{per unit of mass} ${\cal L}$ is a
conserved quantity 

\begin{equation}
 {\cal L} = {\cal L}_\text{ZF}+{\cal L}_\text{PR}=\overline{u}_\phi s + \Omega
s^2 = \text{const.}
 \label{Mom_permass}
\end{equation}
Nondimensionalising this equation by $\Omega r_o^2$ leads to

\begin{equation}
 {\cal L}^* = \text{Ro}\dfrac{s}{r_o}+\dfrac{s^2}{r_o^2}.
 \label{eq:lstar}
\end{equation} 
A spatially
homogeneous ${\cal L}^*$ value would then equal the mass integral of the initial
angular momentum distribution (i.e. a rigid body rotation) given by

%
%

\begin{table}
\caption{$\zeta$ from Eq.~(\ref{eq:mstar1}) and
$\zeta_{TC}$ for different density stratifications.}
\centering
\begin{tabular}{cccc}
  \toprule
  $N_\rho$ & $\zeta$ & $\zeta_{TC}$ \\
  \midrule
 0.01 & 0.470 & 0.616\\
 1    & 0.436 & 0.588 \\
 3    & 0.375 & 0.529 \\
 5    & 0.346 & 0.496 \\
\bottomrule
 \end{tabular}
\label{tab:mstar}
\end{table}

\begin{equation}
 \begin{aligned}
 \zeta =\langle \mom^* \rangle_\rho & = \dfrac{1}{m}\int_V {\cal L}_\text{PR}^*
dm
\\
                          & =
\dfrac{1}{m}\int_0^{2\pi}\int_0^{\pi}\int_{r_i}^{r_o} \left[\dfrac{\rb(r)
s^2}{r_o^2} \right] r^2 \sin\theta\, dr\, d\theta\, d\phi,
 \end{aligned}
 \label{eq:mstar1}
\end{equation}
where $m$ is the total mass of the spherical shell given by $ m= \int_V \rb(r)
dV$. $\zeta$ depends on the background density
stratification and decreases by $25\%$ when the density contrast is increased
from nearly Boussinesq ($N_\rho=10^{-2}$) to $N_\rho=5$ cases (see the
middle column of Tab.~\ref{tab:mstar}). This decrease of $\zeta$
explains the differences in equatorial jet amplitudes in
Fig.~\ref{fig:rossby} via

\begin{equation}
 \text{Ro} = \zeta \dfrac{r_o}{s}-\dfrac{s}{r_o}.
 \label{eq:rotheo}
\end{equation}
Note that \cite{Aurnou07} derived the Boussinesq version of
Eq.~(\ref{eq:rotheo}), in which the background density is homogeneous.

\subsection{Partial mixing of $\mom^*$}

The previous derivation provides an idealised description of the simulation
results at $\rasmid \gg 1$. For instance, the equatorial zonal flows, shown in
Fig.~\ref{fig:rossby}, never reach the maximum
theoretical amplitudes of $\text{Ro}_{e} = -0.528$ for $N_\rho=10^{-2}$ and
$\text{Ro}_{e} =-0.654$ for $N_\rho=5$ predicted by Eq.~(\ref{eq:rotheo}). From
this we suspect that the mixing of $\mom^*$ takes place in only part of the 
spherical shell.

\begin{figure*}[t]
  \centering
  \includegraphics[width=17cm]{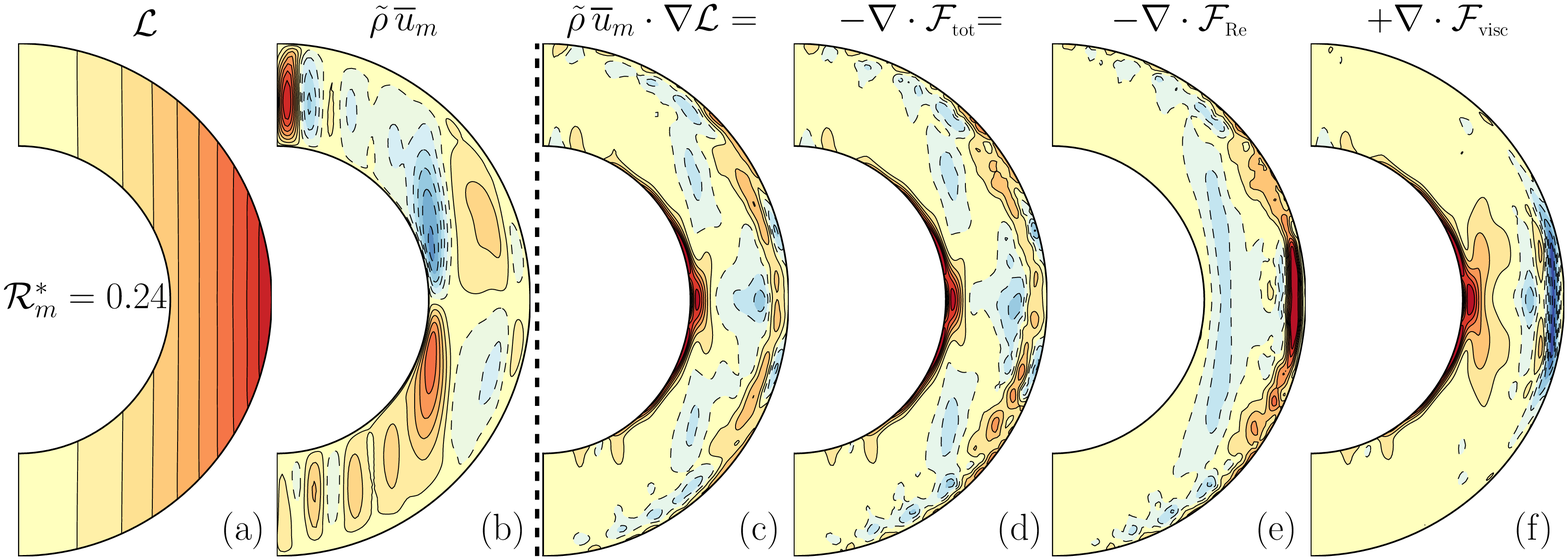}
  \includegraphics[width=17cm]{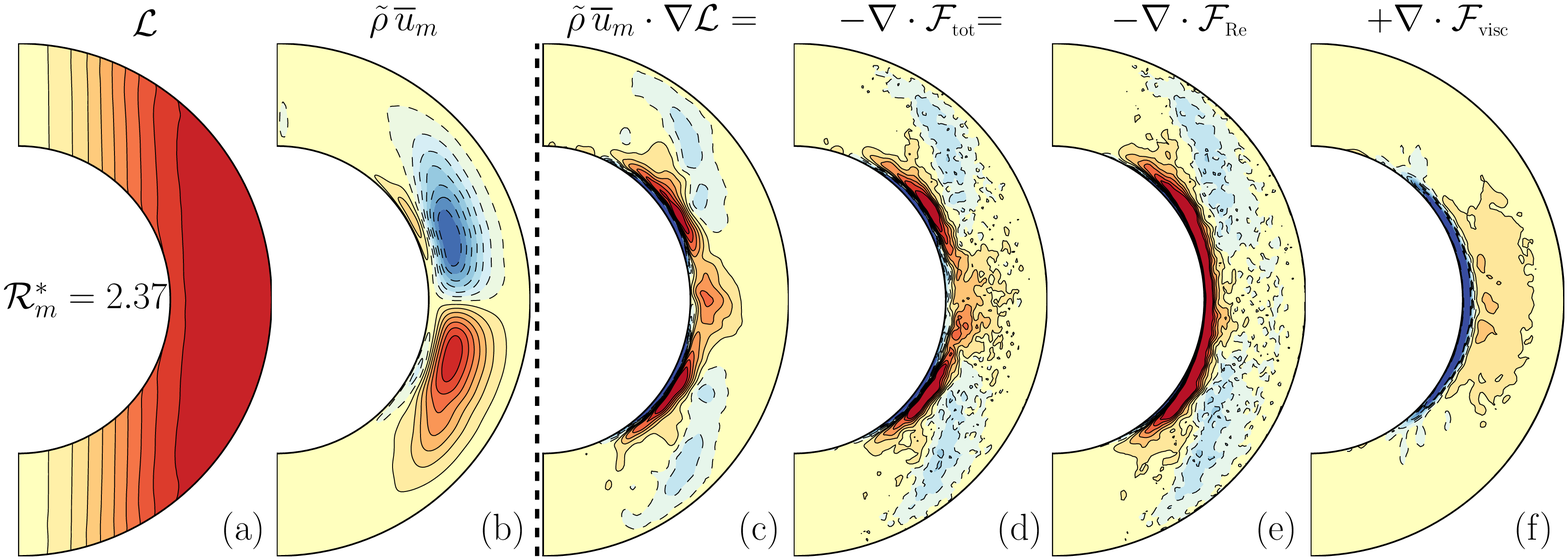}
  \caption{Azimuthal force balance for two numerical models with
$\rasmid=0.24$ 
(upper panels) and $\rasmid =2.37$ (lower panels). Both cases have
$N_\rho=5$ and $\text{E} =10^{-3}$. (a)
Time-averaged angular momentum per unit of mass $\mom^*$.  (b) Time-averaged 
stream-function of the meridional circulation $\rb\,\overline{u}_m$. (c) 
Advection of the angular momentum $\mom^*$ by the mean meridional circulation,
i.e. left-hand side of Eq.~(\ref{eq:mom}). (d) The net axial torque of the
right-hand side of Eq.~(\ref{eq:mom}). (e) The contribution of Reynolds stresses
to this torque. (f) The contribution of viscous stresses to this
axial torque. For each panel, positive (negative) values are rendered in red
(blue). Panels (d-f) share the same contour levels.}
  \label{fig:balN5}
\end{figure*}

We can then make the following \textit{ad-hoc} assumption that $\mom^*$ is
homogenised for cylindrical radii $s>s_\text{mix}$ only. This leads to

\begin{equation}
 \zeta_{s_\text{mix}} =
\dfrac{1}{m(s_\text{mix})} \int_{z(s)} \int_0^{2\pi}\int_{s_\text{mix}}^{r_o}
\left[\dfrac{\rb(r)
s^2}{r_o^2} \right] s\, ds\, d\phi\, dz,
 \label{eq:mstar_cyl}
\end{equation}
where $z(s)=\pm \sqrt{r_o^2-s^2}$ outside the tangent cylinder and $\pm
\lp\sqrt{r_o^2-s^2}-\sqrt{r_i^2-s^2}\rp$ inside. Figure~\ref{fig:mstarNrho}
shows the variation of $\zeta_{s_\text{mix}}$ for different density
stratifications as a function of $s_\text{mix}$.
When the shell is only partially mixed the influence of
the density stratification is gradually reduced and becomes negligible when
$\mom^*$ is homogenised in only a small fraction of the domain. 
Because the tangent cylinder is often considered to present a dynamical barrier
for rotation-dominated convection, \cite{Aurnou07} introduced the respective
values for mixing outside the tangent cylinder only. 
Note, however, that this is a rather arbitrary choice
since there is no expected sharp vorticity step across the tangent cylinder
in  the regime II discussed here. The third column in Tab.~\ref{tab:mstar}
corresponds to these cases (also see the dotted lines in Fig.~\ref{fig:mstarNrho}).
In the Boussinesq limit we recover $\zeta_{TC}=1-\frac{3}{5}(1-\eta^2)$ already
derived by \cite{Aurnou07}.

Figure~\ref{fig:mom} displays time-averaged cylindrically radial profiles of 
$\cal L^*$ and
latitudinal profiles of surface zonal flows for numerical simulations in the
buoyancy-dominated regime (Regime II). For each panel, the grey-shaded area 
shows
the spans between the theoretical $\mom^*$-mixing over the entire shell and
outside the tangent cylinder only. For comparison, a solid body rotation
profile is demarcated by the dashed black lines in the left panels of
Fig.~\ref{fig:mom}.

For $\rasmid \simeq 1$ (dark green lines), the $\mom^*$ profiles are roughly
constant outside the tangent cylinder in the Boussinesq models. From
Tab.~\ref{tab:mstar} and Fig.~\ref{fig:mstarNrho}, we therefore predict a value
of $\text{Ro}_e = \zeta_{s_\text{mix}}-1 \simeq -0.4$ consistent with the
numerical surface value of $\text{Ro}_e = -0.385$ (see Tab.~\ref{tab:results}).
In contrast, $\mom^*$-mixing occurs only in a relatively thin region ($s >
0.7-0.8\,r_o$) in the $N_\rho=5$ case. Predictions for $\text{Ro}_e$ for
this range of mixing depths lie between -0.25 and -0.4, consistent with
the numerical value of $\text{Ro}_e = -0.368$.

In the strongly stratified cases ($N_\rho \geq 3$), the $\mom^*$-mixing moves
deeper when $\rasmid$ is increased. The entire region outside the tangent
cylinder is homogenized (light green line for $N_\rho=3$ and orange line for
$N_\rho=5$) and the surface zonal flows match the theoretical profile derived
before, at least at mid latitudes (i.e. $\theta \lesssim \pm 50^\circ$) once
$\rasmid \simeq 2$. For $\rasmid > 10$, $\mom^*$ is constant for $s >
0.5\,r_o$ and the zonal flow profiles lie roughly midway between the two
theoretical curves (fully-mixed and $s_\text{mix}=s_{TC}$ limits). They reach
stronger amplitude in the $N_\rho=5$ cases due to the larger available $\mom^*$
reservoir (Fig.~\ref{fig:mstarNrho}).

The profiles also show that the $\mom^*$-mixing stops in the highest $\rasmid$
Boussinesq models (dark purple line) and gradually tends toward the solid body
rotation , i.e. $\mom^* = s^2/r_o^2$. As described before, this weakening of the
zonal flow at the highest accessible values of $\rasmid$ appears to define a
third dynamical regime.

\subsection{Azimuthal force balance and meridional flow structure}

In this section, we explore how the zonal flows are maintained in the
$\mom^*$-mixing regime (Regime II in Fig.~\ref{fig:regime}) and the possible
decay of these flows at larger $\rasmid$ (Regime III). To do so, 
we analyse the axisymmetric, azimuthal component of the Navier-Stokes equations
(Eq.~\ref{eq:NS}) following \cite{Miesch11} and \cite{Brun11}:

\begin{equation}
\begin{aligned}
 \rb\dfrac{\partial \overline{u}_\phi}{\partial t} +
\rb\,\overline{\vec{u}}_m\cdot \vec{\nabla}\mom 
 & = -\vec{\nabla}\cdot \vec{{\cal F}}_{\text{tot}}, \\
& = -\vec{\nabla}\cdot
\left[\vec{{\cal F}}_{\text{Re}}-\vec{{\cal F}}_{\text{visc}}\right] , \\
& = -\vec{\nabla}\cdot \left[\rb \,s\, \overline{\vec{u}'u_\phi'} 
-\text{E}\,\rb\, s^2\, \vec{\nabla}\lp\dfrac{\overline{u}_\phi}{s}\rp\right],
\end{aligned}
\label{eq:mom}
\end{equation}
where $\overline{\vec{u}}_m$ is the mean meridional circulation, $\mom
=s\overline{u}_\phi+s^2$ is the angular momentum per unit of mass, 
${\cal F}_{\text{tot}}$ is the angular momentum flux  
associated with Reynolds and viscous stresses, ${\cal F}_{\text{Re}}$ and
${\cal F}_{\text{visc}}$, respectively. Primed quantities correspond to
fluctuations about the axisymmetric mean.
Equation~(\ref{eq:mom}) requires that the advection of axisymmetric
angular momentum by the meridional flow is balanced, on time-average, by the net
axial torques due to Reynolds stresses and viscosity.

Figure~\ref{fig:balN5} shows the angular momentum per unit of mass and the mean
meridional flow, along with the different contributions of this force balance
for two strongly stratified numerical models with $\rasmid = 0.24$ and
$\rasmid=2.37$. In the rotation-dominated case ( $\rasmid=2.37$, upper panels),
the equatorial zonal wind is prograde and $\mom$ gradually increases with
cylindrical radius $s$. In the buoyancy-dominated case ($\rasmid=2.37$, lower
panels), $\mom$ is roughly constant outside the tangent cylinder (see also the
orange curve in the last row of Fig.~\ref{fig:mom}) and is associated with a
strong retrograde equatorial jet which reaches $\text{Ro}_e \simeq -0.5$.

The meridional circulation patterns differ between the two numerical
models shown in Figure~\ref{fig:balN5}. While a relatively small-scale
multicellular meridional circulation structure is observed in the first case,
the meridional flow in the second model is dominated by a pair of large-scale
cells. In the latter, a second pair of weaker counter-cells is discernable close
to the inner boundary. Such a transition between small-scale
multicellular and large-scale single-celled meridional circulation has been
already observed for numerical simulations around  $\rasmid \sim 1$
\citep[e.g.][]{Elliott00,Bessolaz11}. 

\begin{figure}[t]
 \centering
  \includegraphics[width=8.8cm]{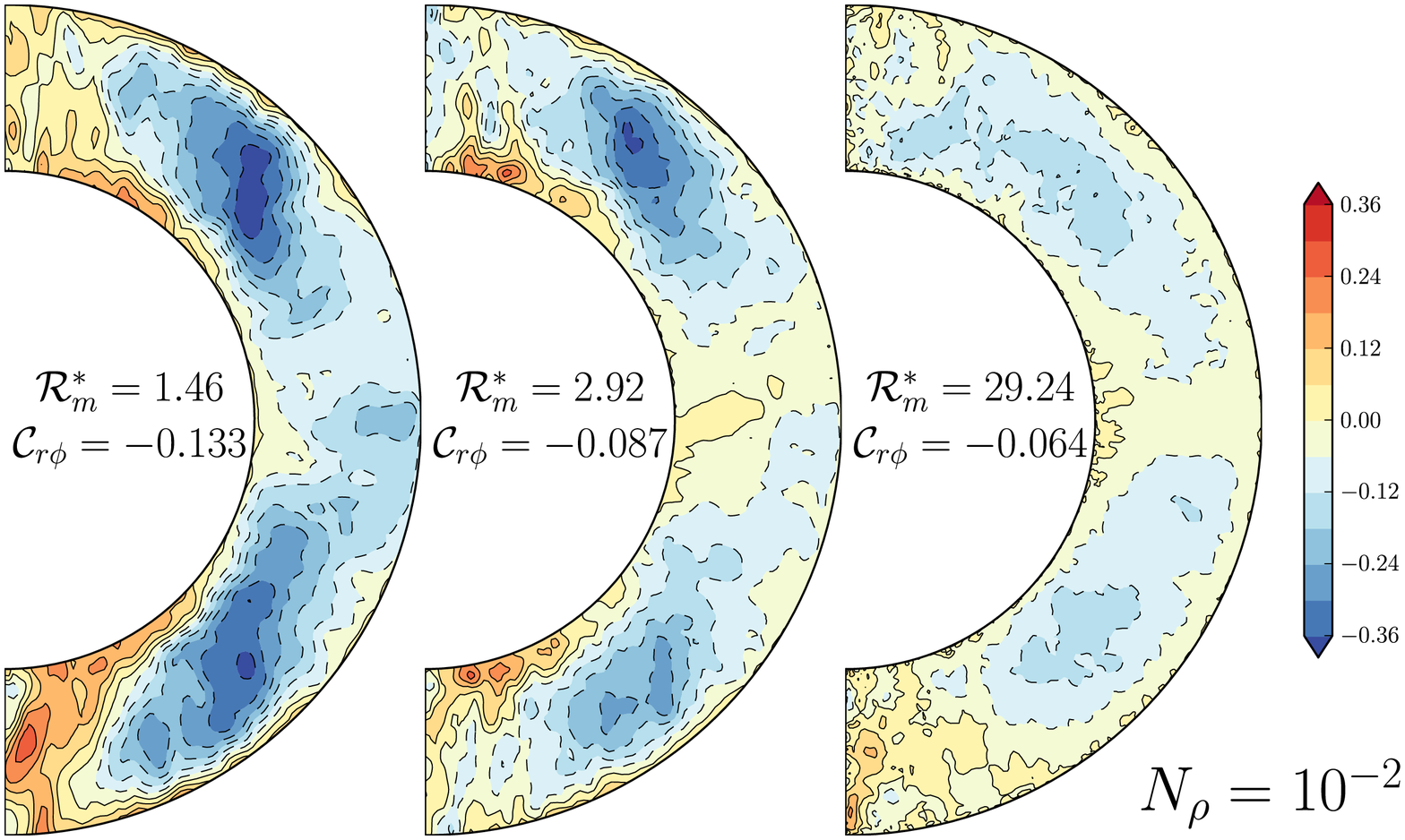}
  \includegraphics[width=8.8cm]{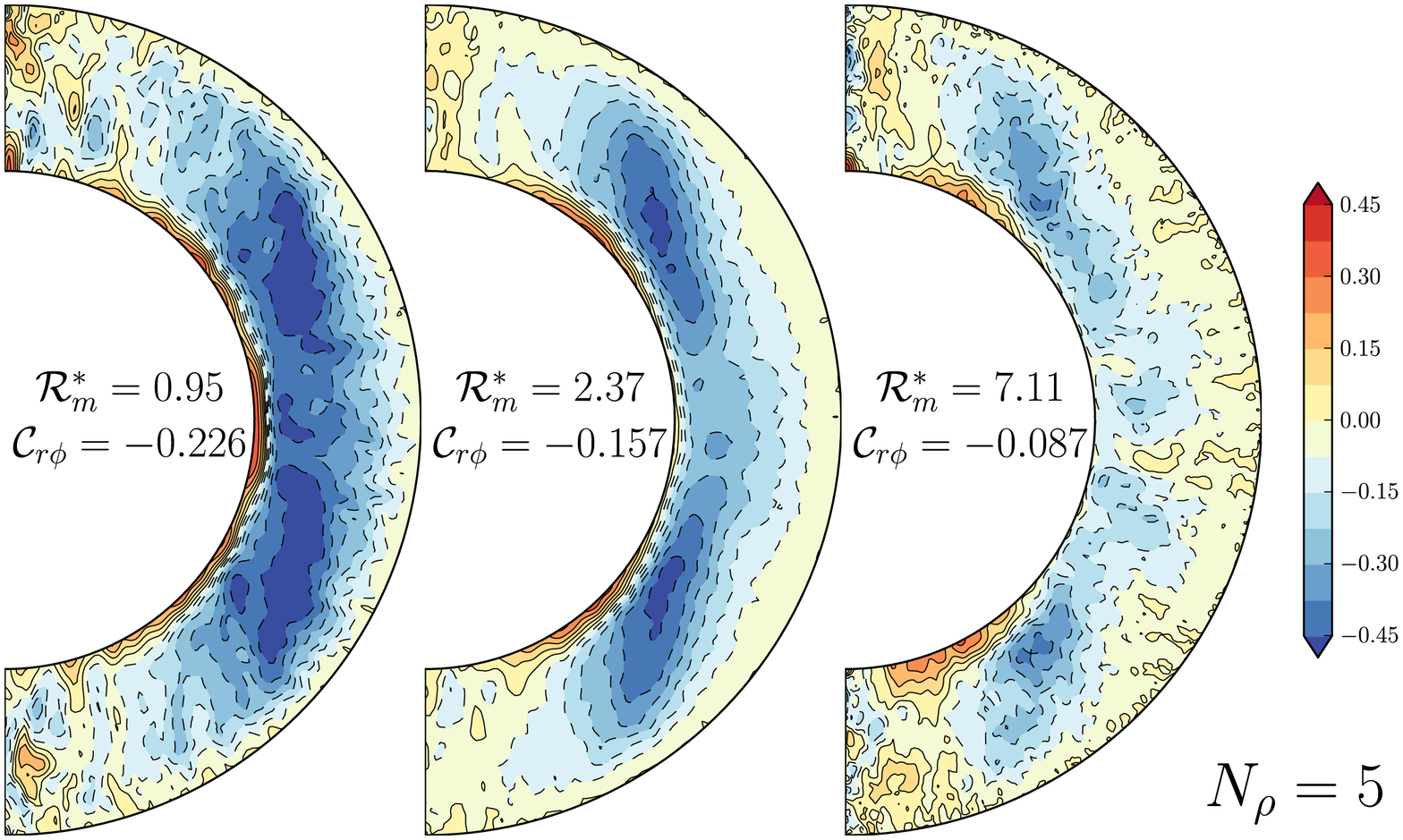}
   \caption{Correlation ${\cal C}_{r\phi}$ for different numerical models in
regime II with $N_\rho=10^{-2}$, $\text{E}=10^{-3}$ (upper panels) and
$N_\rho=5$, $\text{E}=10^{-3}$ (lower panels). The average correlation ${\cal
C}_{r\phi}$ throughout the shell is indicated in the middle of each panel.}
  \label{fig:correl}
\end{figure}

This change of the meridional circulation pattern observed around $\rasmid
\simeq 1$ is reflected in the spatial variations of the force balance
expressed by Eq.~(\ref{eq:mom}). The third and fourth panels of
Fig.~\ref{fig:balN5} check this balance and confirm that the advection of
$\mom$ by the mean meridional flow is indeed balanced, on time average, by the
sum of the torques due to viscous and Reynolds stresses.  
Some small-scale time-dependent features remain
visible in the $\rasmid=2.37$ case but will vanish if a longer
averaging time is considered.
In the $\rasmid = 0.24$ case, viscous stresses (panel f) are significant and
largely, but not perfectly, compensate the Reynolds stresses (panel e). This
results in several sign changes in the axial torque (panel d) that are mirrored
by the complex meridional flow structure \citep[see also][]{Augustson12}. This
force balance is somewhat different in the $\rasmid=2.37$ model in which the
simple large-scale Reynolds stresses dominate the force balance and drive the
dominant pair of meridional circulation cells. The viscous contribution becomes
significant only near the inner boundary and results in the second weaker
pair of cells visible in Fig.~\ref{fig:balN5}b. The presence of these
weak counter cells might however be sensitive to the type of mechanical and
thermal boundary conditions employed in our models \citep[see
also][]{Miesch08}.

\subsection{Towards a possible third regime for $\rasmid \gg 1$}

\begin{figure}
 \centering
  \includegraphics[width=8.8cm]{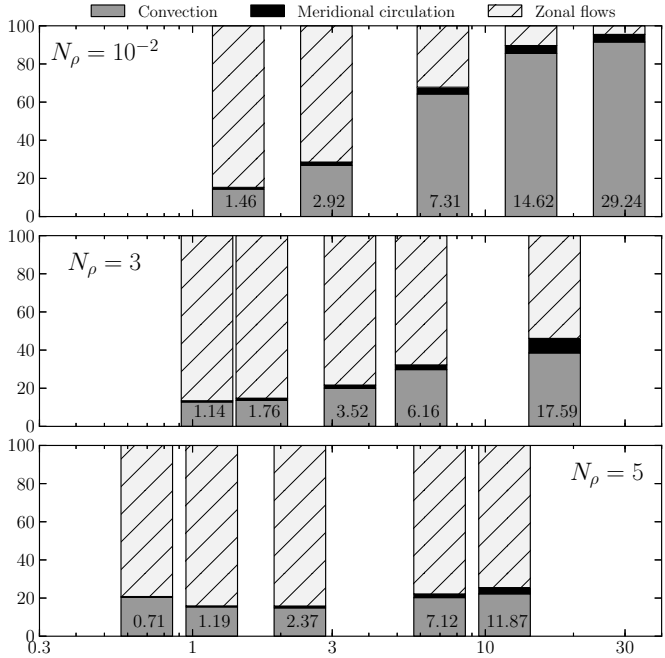}
   \caption{Time-averaged kinetic energy budget for the numerical simulations
displayed in Fig.~\ref{fig:mom}. Dark-grey area correspond to the non-zonal
kinetic energy, black area to the kinetic energy contained in the axisymmetric
poloidal flows (i.e. meridional circulation), and hatched light grey area to the
energy in axisymmetric toroidal flow.}
  \label{fig:budget}
\end{figure}

Figure~\ref{fig:balN5} shows that the Reynolds stresses become the dominant
contribution to the net axial torque when $\rasmid \gtrsim 1$.
The force balance~(\ref{eq:mom}) can thus be approximated in the $\rasmid
\gg 1$ regime by

\begin{equation}
 \rb\,\overline{u}_m\cdot \vec{\nabla} \mom \simeq
-\vec{\nabla}\cdot
\vec{{\cal F}}_{\text{Re}} \simeq - \vec{\nabla}\cdot \left[\rb s\,
\overline{\vec{u}'u_\phi'} \right],
\label{eq:bal_short}
\end{equation}
since viscosity does not play a significant role here.
This balance shows that Reynolds stresses rely on the correlations between the
meridional and the longitudinal components of the convective flow
$\overline{u_r'u_\phi'}$ and $\overline{u_\theta' u_\phi'}$
\citep[e.g.][]{Rudiger89,Kapyla11a}. In the following, we focus on the
change in $\overline{u_r'u_\phi'}$ only and quantify the correlation by

\begin{equation}
 {\cal C}_{r\phi} =
\dfrac{\overline{u_r'u_\phi'}}{\lp\,\overline{u_r'^2}\,\overline{
u_\phi'^2}\,\rp^{1/2}},
\end{equation}
where the overbars correspond to an azimuthal average.
Figure~\ref{fig:correl} shows the variation of ${\cal C}_{r\phi}$ when 
$\rasmid$ is increased in Boussinesq (upper panels) and in strongly
stratified models (lower panels). For $\rasmid \simeq 1$, the
correlations are significant and strong negative Reynolds stresses are
maintained. However, ${\cal C}_{r\phi}$ gradually decreases in the more
supercritical cases. An increase of the Rayleigh number indeed goes along with
stronger turbulent velocities and smaller typical flow lengthscales, leading to
a gradual decrease of the turnover timescale of convection. Small-scale eddies
are not influenced by rotation anymore since their lifetime becomes
significantly smaller than the rotation period \citep[see
also][for the rotation-dominated cases]{Gastine12}.
This loss of coherence results in a gradual decrease of the Reynolds
stresses correlations needed to maintain the mean flows
\citep{Brummell98,Miesch00,Kapyla11a}. A reduction of the mean
flows amplitude is therefore anticipated when the motions become
more turbulent.

Figure~\ref{fig:budget} shows the energy distribution for numerical
models in the buoyancy-dominated regime (see Fig.~\ref{fig:mom} for the
corresponding zonal flows). Zonal flows dominate the energy budget
for $\rasmid \simeq 1$ (80-90\%). In Boussinesq models, the contribution of
the turbulent flows then increases rapidly to overwhelm the
energy distribution for $\rasmid > 10$. This increase is more gradual for
the $N_\rho=3$ cases and it even seems to level around $20\%$ in the
$N_\rho=5$ models. This reflects the density-dependent evolution of
$E_\text{kin}/E_\text{nz}$ in the $\rasmid \gg 1$ regime visible in
Fig.~\ref{fig:rossby}b.

This decorrelation defines a gradual transition towards the possible
regime III displayed in Fig.~\ref{fig:regime}. When the degree of turbulence
increases and
the small-scale motions dominate the energy budget, a decrease
of zonal flows amplitudes is observed in Boussinesq and $N_\rho=1$ models due to
the gradual loss of correlation of the convective flow. 
However, this scenario still needs to be confirmed in strongly stratified
models: the presently accessible range of $\rasmid$ allows for only weakly
turbulent convective motions in our $N_{\rho} \geq 3$ simulations (see
Tab.~\ref{tab:results}).

\section{A transitional regime in anelastic models}
\label{sec:dimple}

As mentioned in section~\ref{sec:model}, $\ras$ is a radially-dependent quantity
that can vary across the fluid shell  by several orders of magnitude in
the strongly stratified models. This can lead to different dynamical
regimes close to the outer boundary and in the fluid shell's
deeper interior.

\subsection{Convective flows in the transitional regime}

\begin{figure}[h]
 \centering
  \includegraphics[width=8.8cm]{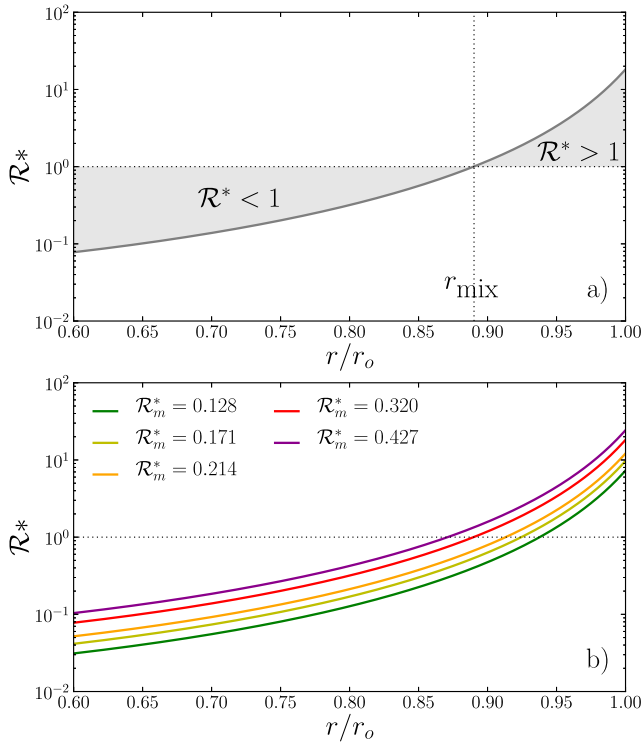}
   \caption{(a) Sketch of the radial profile of $\ras$ in strongly
stratified models. $r_\text{mix}$ corresponds to the radius at which ${\cal
R}^*=1$. It marks the tentative limit between the buoyancy-dominated region
($\ras > 1$) and the rotation-dominated region ($\ras < 1$), both
emphasised by a grey-shaded area. (b) Radial profile of $\ras$ for
various numerical simulations with $N_\rho=5$ and $\text{E}=3\times 10^{-4}$.}
  \label{fig:rastar}
\end{figure}

\begin{figure}
 \centering
  \includegraphics[width=8.8cm]{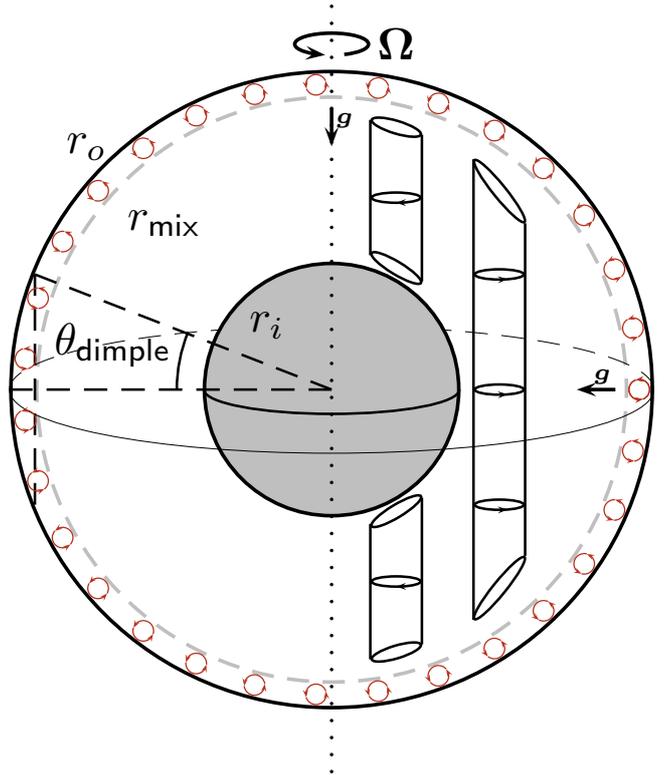}
   \caption{Sketch illustrating the transitional regime in strongly stratified
models.}
  \label{fig:sketch}
\end{figure}

Figure~\ref{fig:rastar} shows the variation of $\ras(r)$ for different
numerical models with $N_\rho=5$. The spherical shell can be separated in two
distinct layers: $\ras > 1$ close to the outer
boundary where buoyancy
effects become large and $\ras < 1$ in the deep interior where rotation
can still dominate the force balance. As $\rasmid$ increases, the
buoyancy-dominated region grows inward.
The radius $r_\text{mix}$ is defined by the radius at which $\ras=1$
approximately separating the buoyancy-dominated region from the
rotation-dominated inner region. Figure~\ref{fig:sketch} sketchs this
\emph{transitional regime} in which columnar convection in
the deep interior exists contemporaneously with three-dimensional convective
structures close to the outer boundary.

Figure~\ref{fig:vortz}, showing equatorial cuts of $\omega_z$ for numerical
models with $N_\rho=5$ and three different Rayleigh numbers, confirms that
$r_\text{mix}$ indeed coincides with a dynamical regime boundary.
In the inner part, the convective flow is dominated by
convective columns tilted in the prograde direction (positive vorticity
dominates) that maintain Reynolds stresses. Beyond $r_\text{mix}$, the
convective flow seems to be more radially-oriented with no preferred sign of
vorticity anymore. Some strongly-stratified simulations discussed in
our previous study already show the very beginning of this transitional regime
\citep[see Fig.~14 in][]{Gastine12}.

\subsection{Zonal flows in the transitional regime}

The transition to regime II happens when $r_\text{mix}$ reaches
approximately mid-depth, where an abrupt transition to a retrograde equatorial
jet takes place. The parameter space, where the transitional regime with two
distinct types of convection co-exists, is therefore quite narrow. 
Strong stratification is required and $\rasmid$ has to be neither too small nor
too large, typically in the $\rasmid \simeq 0.1-0.5$ range.  

\begin{figure*}
 \centering
  \includegraphics[width=17cm]{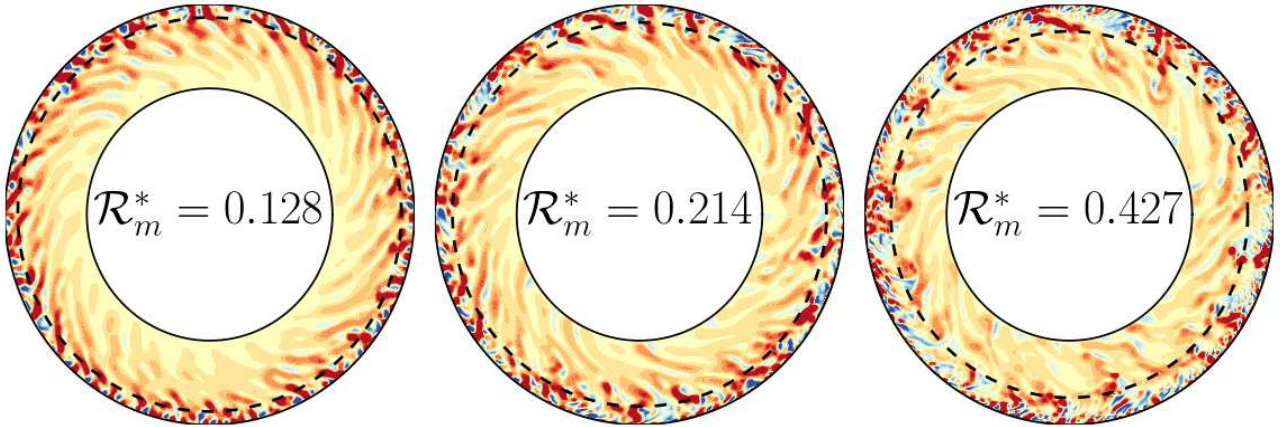}
  \caption{Vorticity along the axis of rotation $\omega_z =
(\vec{\nabla}\times\vec{u})_z$ displayed in the equatorial
plane for three numerical models with $N_\rho = 5$ and $\text{E}=3\times
10^{-4}$. Red (blue)
correspond to positive (negative) values. The dashed black circles correspond to
$r_\text{mix}$ values defined in Fig.~\ref{fig:rastar}.}
  \label{fig:vortz}
\end{figure*}

\begin{figure}[h]
 \centering
  \includegraphics[width=8.8cm]{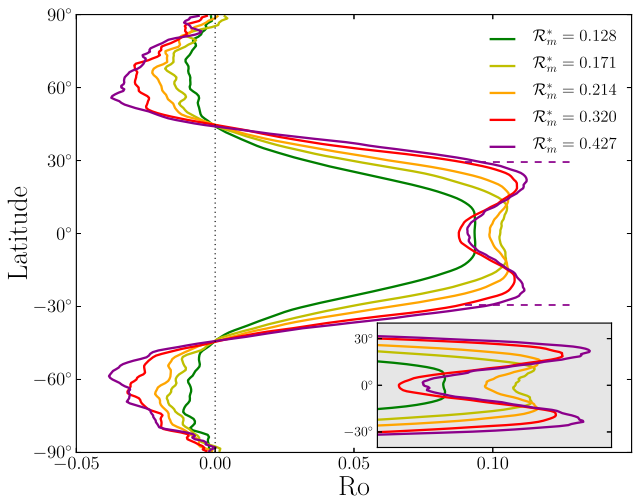}
   \caption{Time-averaged surface zonal flows as a function
of latitude for various numerical simulations with $N_\rho=5$ and 
$\text{E}=3\times 10^{-4}$. The corresponding $\ras(r)$
profiles are given in Fig.~\ref{fig:rastar}. The two magenta lines
correspond to the theoretical width of the dimple obtained from 
Eq.~(\ref{eq:dimple}) for the $\rasmid=0.427$ case.}
  \label{fig:dimple}
\end{figure}

Figure~\ref{fig:dimple} shows the surface zonal flows for numerical models that
lie in this specific range of parameters. Typical for numerical simulations at
moderate Ekman numbers (here $\text{E}=3\times 10^{-4}$) that are still
in the $\rasmid < 1$ regime, a large prograde equatorial jet is flanked by
two weaker retrograde jets at higher latitudes ($\pm 60^\circ$). Due to the
aspect ratio of the spherical shell considered in this study ($\eta=0.6$), the
prograde equatorial wind extends up to mid latitudes of $\pm 50^\circ$. For
$\rasmid = 0.128$ (green line), the zonal wind maximum
is reached at the equator, similarly to Boussinesq models
\citep[e.g.][]{Heimpel05}. However, when increasing $\rasmid$, the center
of the main equatorial jet decreases until eventually a dimple appears flanked
by two maxima. The width of the dimple further grows with $\rasmid$ while the
central flow amplitude decreases reaching approximately $\pm 25^\circ$ latitude
and 20\% of the maximum zonal wind amplitude just before the transition to 
regime II (see the inset in Fig.~\ref{fig:dimple}).
Such a pronounced dimple has also been observed in the anelastic models by
\cite{Kaspi09} who also employed strong density contrasts (see Fig.~13 in their
study). 

Following the sketch displayed in Fig.~\ref{fig:sketch}, this dimple can
be directly related to the transition between rotation-dominated and
buoyancy-dominated regions. As already shown in Fig.~\ref{fig:vortz},
$r_\text{mix}$ separates these two regions
and thus allows us to roughly estimate the typical latitudinal extent of the
dimple via a simple geometrical relation (see Fig.~\ref{fig:sketch})

\begin{equation}
 \cos \theta_\text{dimple} \sim \dfrac{r_\text{mix}}{r_o}.
\label{eq:dimple}
\end{equation}
This simple expression thus relates the latitudinal extent of the observed
surface zonal flows to the transition radius between the two distinct internal
dynamical regimes. Although this expression is rather crude in predicting the 
exact width of the dimple (see the two magenta lines in
Fig.~\ref{fig:dimple}), it can be used as an order of magnitude estimate of its
latitudinal extent.

\subsection{A dimple in the equatorial zonal band of Jupiter}

\begin{figure}
 \centering
  \includegraphics[width=8.8cm]{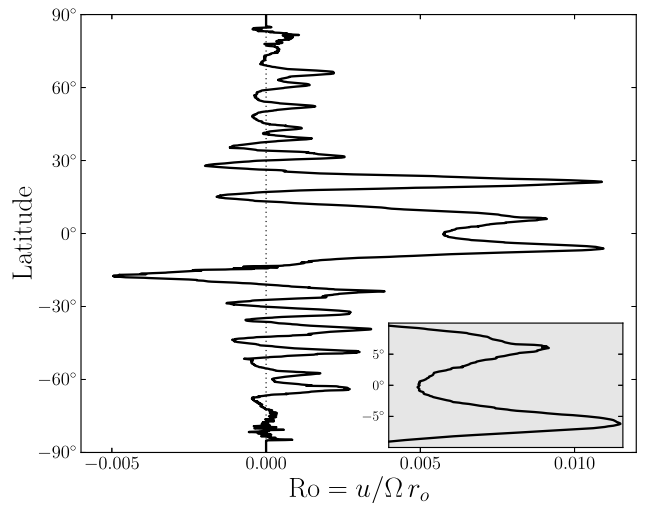}
   \caption{Observed surface zonal winds on Jupiter. Velocities  are given in 
Rossby number units. The data are adapted from \cite{Porco03} and
\cite{Vasavada05}.}
  \label{fig:jupiter}
\end{figure}

A similar dimple exists on Jupiter (Fig.~\ref{fig:jupiter}) and extends roughly
between $\pm 7^\circ$ latitude. The amplitude of the equatorial wind decreases
by roughly 30\% at the equator. Is Jupiter's dimple the surface expression
of an internal regime transition? From Eq.~(\ref{eq:dimple}) with
$\theta_\text{dimple} \sim 7^\circ$, we can speculate that the mixing radius
$r_\text{mix}$ would lie around 500 km below the 1 bar level.

To evaluate the plausibility of this scenario, we try to establish the value
$\ras_\text{surf}$ of Jupiter using the scaling laws by
\cite{Christensen02,Christensen06} and \cite{Gastine12}\footnote{Note that 
using $\text{Nu} \sim \text{Ra}^{1/3}$ as suggested by \cite{King12} leads to
very similar values of $\ras_\text{surf}$.}, which relates the modified Nusselt
number to the flux-based Rayleigh number $\text{Ra}_q^* = \alpha g q/\rho c_p
\Omega^3 d^2$ via $\text{Nu}^* = 0.076\,{\text{Ra}_q^*}^{0.53}$.
$\ras_\text{surf}$ can then directly be obtained with $\ras_\text{surf} =
\text{Ra}_q^*/\text{Nu}^*$. Taking an internal heat flux of
$5.5\,\text{W.m}^{-2}$, $\alpha=6\times 10^{-3}\,\text{K}^{-1}$,
$\Omega=1.75\times 10^{-4}\,\text{s}^{-1}$, $c_p=1.2\times
10^{4}\,\text{J.kg}^{-1}\text{K}^{-1}$, a gravity $g=25\,\text{m.s}^{-2}$, a
density $\rho=0.2\,\text{kg.m}^{-3}$  and a lengthscale
$d=0.04\,r_\text{jup}=2.8\times 10^6\,\text{m}$ \citep[values taken
from][]{French12}, we derive a surface value $\ras_\text{surf} \sim
5\times 10^{-2}$.

The observed near-surface shear layer on the Sun is also thought to be due to 
a
similar dynamical transition when $\ras$ crosses unity \citep[see][]{Miesch11}.
As suggested by these authors, another way to evaluate the impact of rotation
on convection from observable quantities is to estimate the surface convective
Rossby number using $\text{Ro}_c = (\Omega \,\tau_c)^{-1}$, where $\tau_c$ is
the convective turnover time \citep[see also section 4.2 in][]{Gastine12}.
Using $\tau_c=H_\rho/u_\text{conv}$ with
$H_\rho \sim 5\times 10^{4}\,\text{m}$ from the models by \cite{Nettelmann12}
and $u_\text{conv} \sim 5\,\text{m.s}^{-1}$ from observations by \cite{Salyk06},
we obtain $\tau_c \sim 10^{4}\,s$, which leads to $\text{Ro}_c \sim 0.5$ 
(i.e. $\ras_\text{surf} \sim (\text{Ro}_c)^2 \sim 0.25$) at Jupiter's surface,
a factor of five larger than the scaling law prediction.

Concerning the uncertainties in our simulations and in our knowledge
of Jupiter's dynamics, as well as a possible influence of boundary conditions on
scaling laws, this latter value is quite close to the expected value of
$\ras_\text{surf} >1$ required to form a dimple in the transitional regime.

\section{Conclusion}
\label{sec:conclu}

We have investigated the transition
between the rotation-dominated and the buoyancy-dominated regimes in rotating
spherical shells with different density contrasts, extending the previous
Boussinesq study by \cite{Aurnou07}. Following \cite{Glatz1} and
\cite{Jones09}, we have employed the anelastic approximation to filter out fast
acoustic waves and the related short time steps. Exploring moderate
Ekman numbers ($\text{E} = 10^{-3}- 3\times 10^{-4}$) allowed us to raise the
Rayleigh number into the buoyancy-dominated regime (characterised here by
$\rasmid > 1$) and to study zonal flows in a broad parameter range.
We highlight our main findings:

\begin{itemize}
 \item When gradually increasing $\rasmid$, the convective flows change from
geostrophic columnar convection when rotation dominates the force balance to
three-dimensional turbulent motions when buoyancy becomes significant. In the
stratified cases, the latter is characterised by a pronounced asymmetry between
broad upwellings and narrow downwellings. This change in the convective
features is accompanied by a sharp transition in the zonal flow regime. The
equatorial zonal jet reverses its direction at $\rasmid\simeq 1$, independently
of the background density contrast and the Ekman number.

 \item In the rotation-dominated regime (i.e. $\rasmid \ll 1$), a combination
of quasi-geostrophic columns, density stratification effects, and
boundary curvature lead to Reynolds stresses
(i.e. the correlation between the cylindrically radial and the 
azimuthal components of the convective flow) that maintain a prograde equatorial
zonal flow. The zonal flow amplitude is relatively independent of the density 
stratification \citep[see also][]{Gastine12}.

 \item In the buoyancy-dominated regime (i.e. $\rasmid \gg 1$), convection
homogenises the angular momentum per unit of mass which leads to a
strong retrograde equatorial zonal flow flanked by prograde winds at higher
latitudes. The maximum zonal flow amplitude now increases with
density stratification. As already mentioned by \cite{Aurnou07}, these zonal
flow patterns are reminiscent to those observed on Uranus and Neptune, though it
remains uncertain whether convective driving in the ice giants indeed reaches
$\rasmid > 1$.

\item Our simulations suggest the possible existence of a third regime
where the mean zonal flows are negligible and three dimensional
turbulent convection strongly dominates. The timescale of the small-scale
convective motions is much shorter than the rotation period which therefore
cease to play a role in organising large-scale flow. The transition to the third
regime seems to depend on the density stratification and has not been reached
for $N_\rho \geq 3$.

 \item For strongly stratified models in the $\rasmid \simeq 0.1-0.5$ range,
both the dynamical regimes I and II can be present in the spherical shell.
Close to the outer boundary, buoyancy dominates and leads to more turbulent
flows within a thin outer layer. In the deep interior,
rotation still dominates, and columnar convection drive the typical zonal flow
structure. The turbulent outer layer reduces the zonal flow amplitude in the
center of the equatorial jet. This leads to a dimple similar to the one
observed on Jupiter. Its width suggests that a transition between the two
dynamical regimes may occur at a depth of 500 km below the 1 bar level in
Jupiter. Estimate based on the turnover timescale of convection suggests 
$\ras_\text{surf}\sim 0.25$, just on the low side of the required
value.

\item This parameter study on the different zonal flows regimes in rotating
anelastic spherical shells has also some possible stellar applications as it
helps to clarify the transition between the solar-like differential rotation
(i.e. prograde) and the so-called \emph{anti-solar} differential rotation (i.e.
retrograde) observed in a number of stellar convection zone models
\citep[e.g.][]{Brun02,Miesch05,Bessolaz11,Kapyla11}.

\end{itemize}

%

\section*{Acknowledgements}

All the computations have been carried out on the GWDG computer facilities in
G\"ottingen. TG is supported by the Special Priority Program 1488
(PlanetMag, \url{http://www.planetmag.de}) of the German Science Foundation.
JMA gratefully acknowledges the financial support of the US NSF Geophysics
Program.

\bibliographystyle{model2-names}

\appendix

\section{Results table}

{\footnotesize
\onecolumn
\begin{longtable}{ccccccccc}
\caption{Results table.}\\
\toprule
Ekman & $N_\rho$ & ${\cal R}^*_m$ & $\text{Ra}/\text{Ra}_c$ & $\text{Re}_\text{zon}$ & $\text{Re}_\text{mer}$ & $\text{Re}'$ & $\text{Ro}_e$ & Nu \\
\midrule
\endfirsthead
\caption{Continued.}\\
\toprule
Ekman & $N_\rho$ & ${\cal R}^*_m$ & $\text{Ra}/\text{Ra}_c$ & $\text{Re}_\text{zon}$ & $\text{Re}_\text{mer}$ & $\text{Re}'$ & $\text{Ro}_e$ & Nu \\
\midrule
\endhead
\bottomrule
\endfoot
$10^{-3}$ & 0.01 & $2.92 \times 10^{-2}$ & 1.4 & 3.2 & 0.0 & 5.3 & $1.85\times 10^{-3}$ & 1.06\\
$10^{-3}$ & 0.01 & $4.39 \times 10^{-2}$ & 2.1 & 7.0 & 0.1 & 7.7 & $3.64\times 10^{-3}$ & 1.11\\
$10^{-3}$ & 0.01 & $7.31 \times 10^{-2}$ & 3.5 & 14.7 & 0.3 & 12.6 & $6.84\times 10^{-3}$ & 1.21\\
$10^{-3}$ & 0.01 & $1.17 \times 10^{-1}$ & 5.7 & 29.3 & 0.8 & 20.8 & $1.62\times 10^{-2}$ & 1.44\\
$10^{-3}$ & 0.01 & $1.46 \times 10^{-1}$ & 7.1 & 43.4 & 1.1 & 26.4 & $2.55\times 10^{-2}$ & 1.60\\
$10^{-3}$ & 0.01 & $2.19 \times 10^{-1}$ & 10.6 & 77.6 & 3.0 & 41.6 & $4.59\times 10^{-2}$ & 2.18\\
$10^{-3}$ & 0.01 & $2.92 \times 10^{-1}$ & 14.2 & 107.5 & 5.3 & 57.6 & $6.32\times 10^{-2}$ & 2.88\\
$10^{-3}$ & 0.01 & $4.39 \times 10^{-1}$ & 21.2 & 145.4 & 10.7 & 90.0 & $8.32\times 10^{-2}$ & 4.22\\
$10^{-3}$ & 0.01 & $5.85 \times 10^{-1}$ & 28.3 & 150.9 & 11.6 & 121.1 & $1.05\times 10^{-1}$ & 5.36\\
$10^{-3}$ & 0.01 & $7.31 \times 10^{-1}$ & 35.4 & 154.1 & 14.9 & 149.6 & $1.08\times 10^{-1}$ & 6.46\\
$10^{-3}$ & 0.01 & $8.77 \times 10^{-1}$ & 42.5 & 140.3 & 18.1 & 178.7 & $1.05\times 10^{-1}$ & 7.48\\
$10^{-3}$ & 0.01 & 1.02 & 49.6 & 127.8 & 21.0 & 205.5 & $1.01\times 10^{-1}$ & 8.37\\
$10^{-3}$ & 0.01 & 1.17 & 56.6 & 511.1 & 51.5 & 201.5 & $-3.73\times 10^{-1}$ & 8.32\\
$10^{-3}$ & 0.01 & 1.32 & 63.7 & 538.2 & 57.1 & 215.4 & $-3.82\times 10^{-1}$ & 9.16\\
$10^{-3}$ & 0.01 & 1.46 & 70.8 & 561.3 & 58.6 & 231.2 & $-3.85\times 10^{-1}$ & 9.90\\
$10^{-3}$ & 0.01 & 2.92 & 141.6 & 614.2 & 94.6 & 376.6 & $-4.08\times 10^{-1}$ & 15.10\\
$10^{-3}$ & 0.01 & 4.39 & 212.4 & 548.8 & 114.5 & 483.6 & $-4.06\times 10^{-1}$ & 18.70\\
$10^{-3}$ & 0.01 & 5.85 & 283.2 & 471.6 & 142.8 & 583.5 & $-3.88\times 10^{-1}$ & 21.70\\
$10^{-3}$ & 0.01 & 7.31 & 354.0 & 460.6 & 153.3 & 650.5 & $-3.57\times 10^{-1}$ & 23.00\\
$10^{-3}$ & 0.01 & $1.02 \times 10^{1}$ & 495.7 & 432.1 & 191.6 & 802.9 & $-3.03\times 10^{-1}$ & 27.00\\
$10^{-3}$ & 0.01 & $1.46 \times 10^{1}$ & 708.1 & 351.7 & 220.9 & 1010.5 & $-2.64\times 10^{-1}$ & 32.10\\
$10^{-3}$ & 0.01 & $2.92 \times 10^{1}$ & 1416.2 & 333.4 & 313.6 & 1501.5 & $-1.79\times 10^{-1}$ & 44.90\\
\midrule
$10^{-3}$ & 1.00 & $5.47 \times 10^{-2}$ & 1.2 & 2.2 & 0.0 & 4.5 & $1.67\times 10^{-3}$ & 1.04\\
$10^{-3}$ & 1.00 & $6.25 \times 10^{-2}$ & 1.4 & 3.8 & 0.0 & 5.5 & $2.72\times 10^{-3}$ & 1.06\\
$10^{-3}$ & 1.00 & $9.38 \times 10^{-2}$ & 2.1 & 11.6 & 0.1 & 9.8 & $8.25\times 10^{-3}$ & 1.14\\
$10^{-3}$ & 1.00 & $1.25 \times 10^{-1}$ & 2.7 & 20.3 & 0.4 & 13.0 & $1.39\times 10^{-2}$ & 1.23\\
$10^{-3}$ & 1.00 & $1.56 \times 10^{-1}$ & 3.4 & 30.1 & 1.1 & 16.5 & $2.31\times 10^{-2}$ & 1.33\\
$10^{-3}$ & 1.00 & $2.34 \times 10^{-1}$ & 5.1 & 58.3 & 4.5 & 27.4 & $3.81\times 10^{-2}$ & 1.75\\
$10^{-3}$ & 1.00 & $3.13 \times 10^{-1}$ & 6.9 & 77.5 & 9.4 & 36.8 & $6.01\times 10^{-2}$ & 2.22\\
$10^{-3}$ & 1.00 & $4.69 \times 10^{-1}$ & 10.3 & 135.4 & 6.6 & 78.1 & $1.07\times 10^{-1}$ & 3.48\\
$10^{-3}$ & 1.00 & $6.25 \times 10^{-1}$ & 13.7 & 160.6 & 10.0 & 107.8 & $1.38\times 10^{-1}$ & 4.67\\
$10^{-3}$ & 1.00 & $7.81 \times 10^{-1}$ & 17.1 & 164.2 & 13.2 & 137.9 & $1.40\times 10^{-1}$ & 5.75\\
$10^{-3}$ & 1.00 & $9.38 \times 10^{-1}$ & 20.6 & 159.7 & 16.4 & 165.8 & $1.31\times 10^{-1}$ & 6.90\\
$10^{-3}$ & 1.00 & 1.25 & 27.4 & 532.9 & 61.3 & 216.7 & $-4.23\times 10^{-1}$ & 9.13\\
$10^{-3}$ & 1.00 & 1.56 & 34.3 & 573.4 & 73.2 & 255.1 & $-4.31\times 10^{-1}$ & 10.80\\
$10^{-3}$ & 1.00 & 3.13 & 68.5 & 636.4 & 114.2 & 400.3 & $-4.32\times 10^{-1}$ & 16.10\\
$10^{-3}$ & 1.00 & 4.69 & 102.8 & 562.3 & 140.0 & 511.0 & $-4.38\times 10^{-1}$ & 20.10\\
$10^{-3}$ & 1.00 & 6.25 & 137.0 & 462.6 & 140.2 & 584.5 & $-4.03\times 10^{-1}$ & 21.80\\
$10^{-3}$ & 1.00 & 9.38 & 205.5 & 514.7 & 214.5 & 753.3 & $-3.89\times 10^{-1}$ & 28.10\\
\midrule
$10^{-3}$ & 3.00 & $7.92 \times 10^{-2}$ & 1.1 & 1.3 & 0.0 & 4.9 & $2.17\times 10^{-3}$ & 1.03\\
$10^{-3}$ & 3.00 & $8.80 \times 10^{-2}$ & 1.2 & 2.8 & 0.1 & 6.2 & $4.57\times 10^{-3}$ & 1.05\\
$10^{-3}$ & 3.00 & $1.32 \times 10^{-1}$ & 1.8 & 13.8 & 0.3 & 12.6 & $2.04\times 10^{-2}$ & 1.13\\
$10^{-3}$ & 3.00 & $1.76 \times 10^{-1}$ & 2.5 & 27.8 & 0.5 & 18.5 & $3.65\times 10^{-2}$ & 1.23\\
$10^{-3}$ & 3.00 & $2.64 \times 10^{-1}$ & 3.7 & 59.7 & 2.1 & 36.0 & $7.10\times 10^{-2}$ & 1.67\\
$10^{-3}$ & 3.00 & $3.52 \times 10^{-1}$ & 4.9 & 87.3 & 3.6 & 51.0 & $1.04\times 10^{-1}$ & 2.06\\
$10^{-3}$ & 3.00 & $4.40 \times 10^{-1}$ & 6.2 & 110.5 & 5.1 & 64.6 & $1.24\times 10^{-1}$ & 2.42\\
$10^{-3}$ & 3.00 & $5.28 \times 10^{-1}$ & 7.4 & 129.8 & 6.3 & 77.3 & $1.45\times 10^{-1}$ & 2.74\\
$10^{-3}$ & 3.00 & $6.16 \times 10^{-1}$ & 8.6 & 142.8 & 7.8 & 90.0 & $1.59\times 10^{-1}$ & 3.07\\
$10^{-3}$ & 3.00 & $7.04 \times 10^{-1}$ & 9.8 & 153.8 & 8.7 & 101.5 & $1.65\times 10^{-1}$ & 3.32\\
$10^{-3}$ & 3.00 & $8.80 \times 10^{-1}$ & 12.3 & 165.9 & 10.9 & 124.7 & $1.66\times 10^{-1}$ & 3.87\\
$10^{-3}$ & 3.00 & 1.14 & 16.0 & 459.3 & 40.6 & 161.6 & $-4.40\times 10^{-1}$ & 5.44\\
$10^{-3}$ & 3.00 & 1.32 & 18.5 & 501.6 & 50.8 & 177.7 & $-4.62\times 10^{-1}$ & 5.98\\
$10^{-3}$ & 3.00 & 1.76 & 24.6 & 565.5 & 70.1 & 217.6 & $-4.99\times 10^{-1}$ & 7.19\\
$10^{-3}$ & 3.00 & 2.64 & 36.9 & 633.4 & 94.7 & 285.1 & $-5.18\times 10^{-1}$ & 9.00\\
$10^{-3}$ & 3.00 & 3.52 & 49.2 & 669.2 & 110.3 & 335.3 & $-5.25\times 10^{-1}$ & 10.30\\
$10^{-3}$ & 3.00 & 6.16 & 86.1 & 682.3 & 153.1 & 441.0 & $-5.39\times 10^{-1}$ & 12.20\\
$10^{-3}$ & 3.00 & $1.76 \times 10^{1}$ & 246.1 & 753.1 & 325.1 & 625.8 & $-5.41\times 10^{-1}$ & 13.80\\
\midrule
$10^{-3}$ & 5.00 & $4.75 \times 10^{-2}$ & 1.2 & 1.5 & 0.1 & 6.7 & $3.79\times 10^{-3}$ & 1.05\\
$10^{-3}$ & 5.00 & $7.12 \times 10^{-2}$ & 1.8 & 6.2 & 0.6 & 11.5 & $1.37\times 10^{-2}$ & 1.11\\
$10^{-3}$ & 5.00 & $9.49 \times 10^{-2}$ & 2.5 & 13.5 & 0.9 & 18.0 & $2.67\times 10^{-2}$ & 1.19\\
$10^{-3}$ & 5.00 & $1.19 \times 10^{-1}$ & 3.1 & 21.2 & 1.6 & 27.8 & $3.53\times 10^{-2}$ & 1.37\\
$10^{-3}$ & 5.00 & $1.42 \times 10^{-1}$ & 3.7 & 28.5 & 2.4 & 34.8 & $4.76\times 10^{-2}$ & 1.49\\
$10^{-3}$ & 5.00 & $1.66 \times 10^{-1}$ & 4.3 & 35.9 & 2.9 & 40.8 & $5.70\times 10^{-2}$ & 1.59\\
$10^{-3}$ & 5.00 & $1.90 \times 10^{-1}$ & 4.9 & 42.6 & 3.1 & 46.2 & $6.70\times 10^{-2}$ & 1.68\\
$10^{-3}$ & 5.00 & $2.37 \times 10^{-1}$ & 6.2 & 54.3 & 3.6 & 56.4 & $7.68\times 10^{-2}$ & 1.84\\
$10^{-3}$ & 5.00 & $3.56 \times 10^{-1}$ & 9.2 & 77.8 & 5.4 & 80.1 & $9.04\times 10^{-2}$ & 2.24\\
$10^{-3}$ & 5.00 & $4.75 \times 10^{-1}$ & 12.3 & 90.5 & 7.4 & 102.9 & $7.82\times 10^{-2}$ & 2.66\\
$10^{-3}$ & 5.00 & $5.93 \times 10^{-1}$ & 15.4 & 84.2 & 9.7 & 127.3 & $4.54\times 10^{-2}$ & 3.07\\
$10^{-3}$ & 5.00 & $7.12 \times 10^{-1}$ & 18.5 & 318.3 & 24.4 & 138.3 & $-3.68\times 10^{-1}$ & 3.65\\
$10^{-3}$ & 5.00 & $8.31 \times 10^{-1}$ & 21.6 & 357.7 & 27.9 & 150.1 & $-3.98\times 10^{-1}$ & 3.97\\
$10^{-3}$ & 5.00 & $9.49 \times 10^{-1}$ & 24.6 & 384.5 & 31.1 & 163.1 & $-4.21\times 10^{-1}$ & 4.26\\
$10^{-3}$ & 5.00 & 1.19 & 30.8 & 437.4 & 37.1 & 184.5 & $-4.52\times 10^{-1}$ & 4.77\\
$10^{-3}$ & 5.00 & 1.42 & 36.9 & 476.9 & 45.0 & 205.3 & $-4.80\times 10^{-1}$ & 5.23\\
$10^{-3}$ & 5.00 & 2.37 & 61.6 & 574.0 & 68.0 & 270.5 & $-5.33\times 10^{-1}$ & 6.03\\
$10^{-3}$ & 5.00 & 4.75 & 123.2 & 662.8 & 101.6 & 370.8 & $-5.54\times 10^{-1}$ & 7.08\\
$10^{-3}$ & 5.00 & 7.12 & 184.7 & 701.6 & 135.6 & 421.7 & $-5.66\times 10^{-1}$ & 7.23\\
$10^{-3}$ & 5.00 & $1.19 \times 10^{1}$ & 307.9 & 749.0 & 188.3 & 488.4 & $-5.80\times 10^{-1}$ & 7.53\\
\midrule
$3\times10^{-4}$ & 0.01 & $7.89 \times 10^{-3}$ & 1.2 & 2.1 & 0.0 & 4.2 & $3.04\times 10^{-4}$ & 1.03\\
$3\times10^{-4}$ & 0.01 & $1.32 \times 10^{-2}$ & 2.0 & 8.5 & 0.1 & 7.9 & $1.17\times 10^{-3}$ & 1.07\\
$3\times10^{-4}$ & 0.01 & $2.63 \times 10^{-2}$ & 3.9 & 25.4 & 0.3 & 15.7 & $3.86\times 10^{-3}$ & 1.16\\
$3\times10^{-4}$ & 0.01 & $6.58 \times 10^{-2}$ & 9.9 & 106.3 & 1.2 & 39.7 & $1.84\times 10^{-2}$ & 1.60\\
$3\times10^{-4}$ & 0.01 & $7.89 \times 10^{-2}$ & 11.8 & 137.8 & 1.8 & 48.6 & $2.48\times 10^{-2}$ & 1.82\\
$3\times10^{-4}$ & 0.01 & $1.05 \times 10^{-1}$ & 15.8 & 200.7 & 3.3 & 69.6 & $3.74\times 10^{-2}$ & 2.40\\
$3\times10^{-4}$ & 0.01 & $1.32 \times 10^{-1}$ & 19.7 & 263.3 & 5.3 & 90.2 & $4.84\times 10^{-2}$ & 2.98\\
$3\times10^{-4}$ & 0.01 & $1.97 \times 10^{-1}$ & 29.6 & 340.7 & 10.4 & 145.1 & $7.10\times 10^{-2}$ & 4.59\\
$3\times10^{-4}$ & 0.01 & $2.63 \times 10^{-1}$ & 39.4 & 404.8 & 15.9 & 194.4 & $8.95\times 10^{-2}$ & 6.03\\
$3\times10^{-4}$ & 0.01 & $3.95 \times 10^{-1}$ & 59.1 & 492.4 & 26.4 & 282.5 & $1.09\times 10^{-1}$ & 8.75\\
$3\times10^{-4}$ & 0.01 & $5.26 \times 10^{-1}$ & 78.8 & 565.8 & 36.7 & 380.0 & $9.26\times 10^{-2}$ & 11.70\\
$3\times10^{-4}$ & 0.01 & $7.89 \times 10^{-1}$ & 118.2 & 615.9 & 53.3 & 571.7 & $6.53\times 10^{-2}$ & 16.70\\
$3\times10^{-4}$ & 0.01 & 1.05 & 157.6 & 2069.2 & 172.7 & 550.7 & $-3.76\times 10^{-1}$ & 19.00\\
$3\times10^{-4}$ & 0.01 & 1.32 & 197.0 & 2172.9 & 181.4 & 656.9 & $-4.11\times 10^{-1}$ & 22.60\\
$3\times10^{-4}$ & 0.01 & 2.63 & 394.1 & 2151.6 & 232.4 & 1093.1 & $-4.25\times 10^{-1}$ & 35.90\\
\midrule
$3\times10^{-4}$ & 1.00 & $2.11 \times 10^{-2}$ & 1.2 & 2.7 & 0.0 & 4.7 & $5.35\times 10^{-4}$ & 1.02\\
$3\times10^{-4}$ & 1.00 & $2.81 \times 10^{-2}$ & 1.5 & 8.9 & 0.0 & 8.2 & $1.73\times 10^{-3}$ & 1.06\\
$3\times10^{-4}$ & 1.00 & $4.22 \times 10^{-2}$ & 2.3 & 25.6 & 0.1 & 14.1 & $5.26\times 10^{-3}$ & 1.13\\
$3\times10^{-4}$ & 1.00 & $5.63 \times 10^{-2}$ & 3.1 & 49.9 & 0.4 & 22.6 & $1.06\times 10^{-2}$ & 1.23\\
$3\times10^{-4}$ & 1.00 & $8.44 \times 10^{-2}$ & 4.6 & 105.8 & 1.2 & 39.0 & $2.55\times 10^{-2}$ & 1.51\\
$3\times10^{-4}$ & 1.00 & $1.13 \times 10^{-1}$ & 6.2 & 162.3 & 2.4 & 57.1 & $4.04\times 10^{-2}$ & 1.91\\
$3\times10^{-4}$ & 1.00 & $1.41 \times 10^{-1}$ & 7.7 & 210.4 & 3.9 & 75.4 & $5.07\times 10^{-2}$ & 2.36\\
$3\times10^{-4}$ & 1.00 & $2.11 \times 10^{-1}$ & 11.6 & 302.5 & 8.8 & 125.7 & $7.39\times 10^{-2}$ & 3.70\\
$3\times10^{-4}$ & 1.00 & $2.81 \times 10^{-1}$ & 15.5 & 379.2 & 13.4 & 173.1 & $1.02\times 10^{-1}$ & 5.09\\
$3\times10^{-4}$ & 1.00 & $5.63 \times 10^{-1}$ & 31.0 & 592.1 & 32.7 & 340.6 & $1.35\times 10^{-1}$ & 10.20\\
$3\times10^{-4}$ & 1.00 & $8.44 \times 10^{-1}$ & 46.5 & 672.2 & 52.9 & 556.8 & $4.46\times 10^{-2}$ & 16.90\\
$3\times10^{-4}$ & 1.00 & 1.13 & 62.0 & 2088.3 & 145.0 & 614.0 & $-3.47\times 10^{-1}$ & 23.10\\
\midrule
$3\times10^{-4}$ & 3.00 & $3.17 \times 10^{-2}$ & 1.1 & 1.2 & 0.0 & 4.8 & $7.32\times 10^{-4}$ & 1.02\\
$3\times10^{-4}$ & 3.00 & $4.75 \times 10^{-2}$ & 1.6 & 15.1 & 0.2 & 12.6 & $6.80\times 10^{-3}$ & 1.08\\
$3\times10^{-4}$ & 3.00 & $6.33 \times 10^{-2}$ & 2.2 & 40.8 & 0.4 & 21.2 & $1.73\times 10^{-2}$ & 1.17\\
$3\times10^{-4}$ & 3.00 & $7.92 \times 10^{-2}$ & 2.7 & 72.6 & 0.7 & 30.0 & $3.00\times 10^{-2}$ & 1.27\\
$3\times10^{-4}$ & 3.00 & $1.19 \times 10^{-1}$ & 4.0 & 152.3 & 2.8 & 60.1 & $5.84\times 10^{-2}$ & 1.82\\
$3\times10^{-4}$ & 3.00 & $1.58 \times 10^{-1}$ & 5.4 & 227.9 & 5.3 & 89.1 & $8.23\times 10^{-2}$ & 2.44\\
$3\times10^{-4}$ & 3.00 & $2.38 \times 10^{-1}$ & 8.1 & 363.9 & 9.7 & 140.9 & $1.27\times 10^{-1}$ & 3.60\\
$3\times10^{-4}$ & 3.00 & $3.17 \times 10^{-1}$ & 10.8 & 469.4 & 13.7 & 187.3 & $1.58\times 10^{-1}$ & 4.66\\
$3\times10^{-4}$ & 3.00 & $4.75 \times 10^{-1}$ & 16.2 & 572.5 & 21.4 & 275.8 & $1.83\times 10^{-1}$ & 6.76\\
$3\times10^{-4}$ & 3.00 & $6.33 \times 10^{-1}$ & 21.5 & 601.7 & 28.9 & 354.1 & $1.52\times 10^{-1}$ & 8.04\\
$3\times10^{-4}$ & 3.00 & $7.92 \times 10^{-1}$ & 26.9 & 513.6 & 40.1 & 492.3 & $2.86\times 10^{-2}$ & 10.60\\
$3\times10^{-4}$ & 3.00 & 1.19 & 40.4 & 2038.5 & 158.1 & 518.3 & $-5.13\times 10^{-1}$ & 16.90\\
\midrule
$3\times10^{-4}$ & 5.00 & $1.50 \times 10^{-2}$ & 1.1 & 0.8 & 0.0 & 5.8 & $8.20\times 10^{-4}$ & 1.03\\
$3\times10^{-4}$ & 5.00 & $2.14 \times 10^{-2}$ & 1.6 & 6.6 & 0.1 & 11.5 & $5.45\times 10^{-3}$ & 1.06\\
$3\times10^{-4}$ & 5.00 & $3.20 \times 10^{-2}$ & 2.4 & 21.3 & 0.7 & 21.3 & $1.54\times 10^{-2}$ & 1.16\\
$3\times10^{-4}$ & 5.00 & $4.27 \times 10^{-2}$ & 3.1 & 41.4 & 1.6 & 34.8 & $2.60\times 10^{-2}$ & 1.35\\
$3\times10^{-4}$ & 5.00 & $6.41 \times 10^{-2}$ & 4.7 & 93.8 & 3.5 & 72.1 & $4.73\times 10^{-2}$ & 2.00\\
$3\times10^{-4}$ & 5.00 & $8.54 \times 10^{-2}$ & 6.3 & 149.9 & 4.9 & 98.1 & $6.71\times 10^{-2}$ & 2.46\\
$3\times10^{-4}$ & 5.00 & $1.28 \times 10^{-1}$ & 9.4 & 249.5 & 8.2 & 144.8 & $9.68\times 10^{-2}$ & 3.27\\
$3\times10^{-4}$ & 5.00 & $1.71 \times 10^{-1}$ & 12.6 & 323.5 & 11.5 & 185.3 & $1.05\times 10^{-1}$ & 3.96\\
$3\times10^{-4}$ & 5.00 & $2.14 \times 10^{-1}$ & 15.7 & 372.0 & 15.0 & 224.7 & $9.74\times 10^{-2}$ & 4.67\\
$3\times10^{-4}$ & 5.00 & $3.20 \times 10^{-1}$ & 23.5 & 448.5 & 20.2 & 295.7 & $8.78\times 10^{-2}$ & 5.53\\
$3\times10^{-4}$ & 5.00 & $4.27 \times 10^{-1}$ & 31.4 & 470.8 & 26.2 & 361.0 & $9.09\times 10^{-2}$ & 6.69\\
$3\times10^{-4}$ & 5.00 & $6.41 \times 10^{-1}$ & 47.1 & 477.2 & 35.6 & 454.6 & $7.71\times 10^{-2}$ & 8.75\\
$3\times10^{-4}$ & 5.00 & $8.54 \times 10^{-1}$ & 62.8 & 1550.8 & 121.8 & 435.8 & $-4.76\times 10^{-1}$ & 9.92\\
\midrule
\label{tab:results}
\end{longtable}
}

\end{document}